\documentclass[11pt,a4paper]{article}
\usepackage[T1]{fontenc}
\usepackage[utf8]{inputenc}
\usepackage{amsmath}
\usepackage{newtxtext,newtxmath}
\usepackage{bm}
\usepackage{graphicx}
\usepackage{booktabs}
\usepackage{multirow}
\usepackage[textwidth=398pt,top=2.5cm,bottom=2.5cm,headheight=14pt]{geometry}
\usepackage[font=small,labelfont=bf,labelsep=period]{caption}
\usepackage[hidelinks]{hyperref}
\usepackage{apacite}
\graphicspath{{figures/}}
\newcommand{\E}{\mathbb{E}}
\newcommand{\CorrespondingEmail}{fventrella37@gmail.com}

\title{Estimating Vertical Ocean Velocity from Horizontal Currents:
Linear and Depth-Conditional Neural-Network Approaches
on a High-Resolution Regional Ocean Model}
\author{Francesco M. Ventrella\textsuperscript{1},
  Crist\'obal L\'opez\textsuperscript{1} and
  Emilio Hern\'andez-Garc\'ia\textsuperscript{1}\\[1.5ex]
  {\small\textsuperscript{1}Instituto de F\'{\i}sica Interdisciplinar y Sistemas
  Complejos (IFISC), CSIC-UIB,}\\
  {\small Campus Universitat de les Illes Balears, E-07122 Palma de Mallorca, Spain}\\[1.5ex]
  {\small Corresponding author: Francesco M. Ventrella, \CorrespondingEmail}}
\date{}

\begin{document}
\maketitle

\begin{abstract}
The vertical component of the ocean velocity, $w$, is several orders of
magnitude smaller than the horizontal components $u$ and $v$, yet it
controls vertical tracer transport, upwelling, and mixed-layer ventilation.
Because $w$ is extremely difficult to measure directly and is normally
obtained only diagnostically from the continuity equation, predicting it
from the horizontal velocity field is a problem of both practical
and physical interest. We address this problem with two data-driven
methods, a physically-motivated linear baseline and a convolutional neural
network conditioned on depth through Feature-wise Linear Modulation
(FiLM). Both methods are applied to a high-resolution regional simulation
of the Coastal and Regional Ocean Community model over the Western
Mediterranean, using a sub-region of the south-west Mediterranean Sea as a
training region and two independent, out-of-sample regions for validation:
the Alboran Sea, with a distinct dynamical regime driven by Atlantic
inflow through the Strait of Gibraltar, and a transition region
between Almer\'{\i}a and Cartagena, between the Alboran Sea and
the open western Mediterranean. Once the FiLM network is
regularized enough to avoid overfitting to a single training domain, we
find that its comparison with the physically interpretable
linear baseline is domain- and depth-dependent rather than uniform: near
the reference depth, essentially all of the achievable out-of-sample
performance on both validation domains is already captured by the linear
model, but at the deepest levels tested the FiLM network outperforms the
linear baseline on the transition domain, whereas on Alboran it does not.
\end{abstract}

\section{Introduction}
\label{sec:intro}

The vertical component of ocean velocity, $w$, plays a central role in
vertical tracer transport, upwelling dynamics \cite{Brink1983}, and
mixed-layer ventilation, particularly at the
submesoscale \cite{McWilliams2016}.
Despite its dynamical importance, $w$ is orders of magnitude smaller than
the horizontal components $u$ and $v$ \cite{HeMahadevan2024,Liang2017}, making direct
measurement extremely difficult: essentially all in-situ and satellite
observing systems resolve the horizontal velocity field (via
altimetry \cite{Fu2010,Morrow2023}, drifters \cite{Lumpkin2017}, or Doppler
techniques \cite{PollardRead1989}) far more directly than they resolve
$w$. The vertical velocity is instead typically inferred diagnostically
from horizontal velocities \cite{Liang2017} or from hydrographic data. Moreover, horizontal currents are
routinely mapped near the surface, by satellite altimetry, surface
drifters and coastal high-frequency radar
networks \cite{PaduanWashburn2013,Roarty2019}, whereas below the surface
they are available only from sparse moorings and profilers. Predicting
$w$ at depth from near-surface horizontal velocity alone is therefore of
direct practical interest.

Machine learning (ML) has, over the last decade, become a standard tool
throughout ocean science, from parameterizing unresolved sub-grid physics
to reconstructing under-observed fields from sparse or indirect
measurements. Recent reviews describe its applications to ocean
observation, theory, and simulation \cite{Sonnewald2021, Lou2023, Kim2022}.
Data-driven approaches have been used to parameterize vertical mixing in
the ocean surface boundary layer \cite{Liang2022}, to estimate mixed-layer
depth from sparse profiles and dense satellite
data \cite{Foster2021,Imchen2025}. In atmospheric modeling, a hybrid
approach in which a learned component corrects a physics-based numerical
model has also been demonstrated \cite{Arcomano2022}. In the ocean, machine
learning has been applied to
reconstruct ocean surface currents directly from more easily observed
quantities \cite{Sinha2021} and to extract mesoscale and submesoscale
surface dynamics from sparse Lagrangian drifter
trajectories \cite{Aksamit2020}, both of which pose the same basic
question addressed here, recovering a dynamically important but
poorly-observed velocity-related quantity from more accessible data. Most
directly relevant to this manuscript, \citeA{HeMahadevan2024} have already used machine learning to
diagnose $w$ from surface data: they trained linear
regression, random-forest, and convolutional neural network (CNN) models to
diagnose the three-dimensional submesoscale $w$ field from surface
velocity, density, and their horizontal gradients in an idealized,
periodic-channel simulation of an upwelling front, finding that a CNN
using only surface fields ($u$, $v$ and density) recovers most of the
achievable performance and is comparatively robust to noisy or coarsened input
data. \citeA{Zhu2023} trained a deep neural network to
reconstruct the three-dimensional $w$ in the upper $500$~m from sea surface
height alone, using as input a patch of about $45\times45$~km around the
target point, and found it more accurate than dynamically based
diagnostic methods. This manuscript asks a closely related question, how well $w$
can be recovered from $u$ and $v$ alone, on a realistic, non-idealized regional simulation with actual
coastline and bathymetry, and evaluates spatial generalization to
independent, dynamically distinct out-of-sample regions rather than a
train/test split drawn from the same simulation ensemble. Beyond
velocity fields, ML regression has been used to bias-correct
satellite-derived sea surface temperature \cite{SauxPicart2018}, to
estimate subsurface temperature from surface
data \cite{Wang2021}, to predict water-mass mixing from hydrographic
tracers \cite{RomeraCastillo2022}, to reconstruct dissolved organic carbon
concentrations \cite{Laine2024}, to estimate deep-ocean particle
flux \cite{Picard2025}, to fill gaps in ocean mooring time
series \cite{Sloyan2023}, and, closer to our own regional focus, to predict
sea surface temperature and marine heatwave occurrence specifically in the
Mediterranean Sea \cite{Bonino2024}. In fluid mechanics proper, a related
precedent is the use of a CNN to reconstruct the (also difficult to
directly measure) skin-friction drag over ocean surface waves from more
accessible flow quantities \cite{Yousefi2024}. In turbulence
research, deep networks have been used to infer an unmeasured velocity
component from the measured ones in rotating turbulent
flows \cite{Li2023} and to reconstruct missing
regions of turbulent fields \cite{Buzzicotti2021}. General treatments
of the mathematical and algorithmic foundations of deep learning used
throughout this manuscript can be found in \citeA{Berlyand2023} and \citeA{Drori2022}.

This manuscript investigates whether $w$ at different depths can be accurately
predicted from $u$ and $v$ alone, taken at a single near-surface level on a small square of
neighboring grid points around the target point (a stencil;
\citeNP{Sinha2021}), using
data-driven approaches ranging from a physically
motivated linear model to a depth-conditional convolutional neural
network. The key scientific questions are: (i) how deep does the
statistical relationship between the horizontal stencil and $w$ remain
informative when a model trained at one location is applied to an
independent, dynamically distinct region; and (ii) whether generalizing
the linear, per-level baseline to an arbitrary depth through a
non-linear, depth-conditional neural network yields a genuine gain in
predictive performance, or whether the added flexibility proves unnecessary once
the physically-motivated linear structure is already exploited.

We use output from the realistic $500$-m-resolution regional simulation
of the Alboran Sea and the adjacent western Mediterranean of
\citeA{Capo2021}, run
with the CROCO (Coastal and Regional Ocean COmmunity)
model \cite{CROCO2022} on a curvilinear $\sigma$-coordinate grid and
building on the ROMS
family of models \cite{ShchepetkinMcWilliams2005,Haidvogel1991}. All models are
trained on a sub-region of the south-west Mediterranean Sea, south
of the islands of Ibiza and Formentera, and validated on the
Alboran Sea, an independent region about $400$~km away with a
distinct dynamical regime dominated by the Atlantic inflow through the
Strait of Gibraltar, providing a stringent spatial out-of-sample test, and on a
second independent region, in the open sea between Almer\'{\i}a and
Cartagena ($36.4$--$37.4^\circ$N, $1.5$--$0.2^\circ$W), which is a
transition region between the Alboran Sea and the open western
Mediterranean (Section~\ref{sec:domains}).

The remainder of the manuscript is organized as follows.
Section~\ref{sec:methods} describes the data, the theoretical background
and the methods, Section~\ref{sec:results} reports the corresponding
results, and Section~\ref{sec:conclusions} discusses their implications and
outlines future extensions. The Supporting Information reports two
further tests: the sensitivity of the models to errors in the input
velocities (Text S1) and their performance when applied to dates from
different times of the year than those used for training (Text S2).

\section{Methods}
\label{sec:methods}

\begin{figure}[!htbp]
  \centering
  \includegraphics{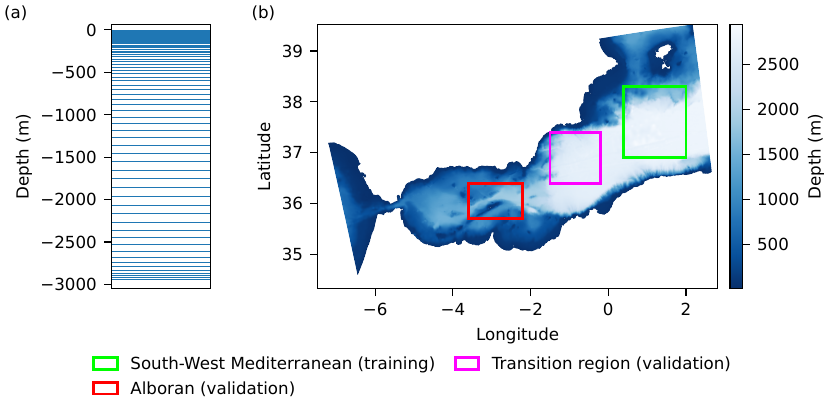}
  \caption{(a) The fixed $z$-levels used throughout this manuscript (70
    levels, from $-2.7$ to $-2935$~m). Spacing is finest near the surface
    and coarsens with depth, following the native $\sigma$-coordinate
    resolution. (b) Training domain (South-West Mediterranean, green) and
    the two independent out-of-sample validation domains (Alboran, red;
    transition region, magenta), overlaid on the CROCO bathymetry used
    throughout this manuscript.}
  \label{fig:zlevels_domains}
\end{figure}

All methods below operate on a single time instant of the
simulation of Cap\'o et al., which has a horizontal resolution of
$500$~m and $70$ $\sigma$-levels and stores its output as $3$-hour
averages \cite{Capo2021}. It is one-way nested in a $1.5$-km simulation of
the western Mediterranean, which provides its initial state and the
temperature, salinity, velocity and sea level at its open boundaries,
where tides are also imposed; at the surface it is forced by hourly
atmospheric fields of the NCEP Climate Forecast System Reanalysis (NCEP-CFSR), through
bulk formulae for the air--sea fluxes \cite{Capo2021}. We use the average centered on 3
February 2011, 19:30~UTC (18:00--21:00~UTC). This snapshot falls in winter: sea-surface temperature is
$\approx14$--$16^\circ$C over the three domains used here, and the
surface boundary layer diagnosed by the model \cite{Capo2021} is a few
tens of meters deep on average and up to about $120$~m. Winter is also the season in which
frontogenesis, and the submesoscale vertical motion associated with it,
is most active in the Alboran Sea \cite{Capo2021}. All results of the main text therefore refer to
winter conditions; whether they carry over to other seasons is tested
in Text S2 of the Supporting Information.
The native model grid is a curvilinear, terrain-following Arakawa
C-grid \cite{ArakawaLamb1977},
on which $u$, $v$, and $w$ are staggered at different horizontal and
vertical points rather than co-located, and whose local grid angle
relative to true north is not spatially constant (it varies between
$7.4^\circ$ and $13.6^\circ$ across the model domain); we retain this native
grid throughout, without rotating $(u,v)$ onto the geographic frame.
Horizontal derivatives are correspondingly taken along the native
curvilinear grid directions $(\xi,\eta)$, and $u$ and $v$ are the
horizontal velocity components of the CROCO model along these two directions. These two
directions are
orthogonal at every grid point, by construction of the model
grid \cite{CROCO2022}, so that $(\xi,\eta)$ is locally a Cartesian frame,
rotated with respect to the geographic one. The horizontal
divergence entering the continuity equation therefore keeps its Cartesian
form, with no cross terms, and reads $\partial u/\partial\xi +
\partial v/\partial\eta$ rather than $\partial u/\partial x+\partial
v/\partial y$.
Before any of the analyses below, $u$, $v$, and $w$ are linearly interpolated onto
a common set of horizontal grid points and a common set of fixed
$z$-levels, so that all three components are evaluated at the same
location in space. This is a convenient adjustment, since the
incompressibility condition $\nabla_{\xi\eta z}\cdot\mathbf{u}=0$, with
$\mathbf{u}=(u,v,w)$, is a local, point-wise relation. These
fixed $z$-levels are not uniformly spaced: as shown in
Figure~\ref{fig:zlevels_domains}a, they are closely packed near the surface and
progressively coarser with depth, matching the vertical resolution of the
native $\sigma$-coordinate grid they are interpolated from. All the
quantities introduced below, i.e.\ the divergence, the vertical correlations
of $w$ and the stencils used as model inputs, are evaluated on these
interpolated fields.

\subsection{Linear structure of the incompressibility equation}
\label{sec:continuity}

Conceptually, the physical starting point of this manuscript is the continuity equation for
an incompressible fluid,
\begin{equation}
  \nabla_{\xi\eta z}\cdot\mathbf{u} = \frac{\partial u}{\partial \xi} + \frac{\partial v}{\partial \eta}
  + \frac{\partial w}{\partial z} = 0 ,
  \label{eq:continuity}
\end{equation}
which relates the vertical velocity $w$ to the horizontal velocities $u,v$
through their spatial derivatives. Here $\xi$ and $\eta$ are the
horizontal grid directions introduced above, $u$ and $v$ the velocity
components along them, and $z$ the vertical coordinate. Crucially, Equation~\ref{eq:continuity} is
linear in $u$, $v$, and $w$: no product or other nonlinear function
of the velocity components appears. Equivalently, at any fixed depth $w$ is
a linear functional of the horizontal divergence field
$\partial u/\partial\xi + \partial v/\partial\eta$ and of its own value at
a reference level. This linear structure is
the physical motivation for the linear regression baseline introduced in
Section~\ref{sec:lr}: rather than assuming any particular finite-difference
form of Equation~\ref{eq:continuity}, we ask whether a general linear map from
a local horizontal-velocity stencil to $w$ can be recovered directly from
data, and, if so, whether it reproduces the divergence operator that
Equation~\ref{eq:continuity} already prescribes.

A caveat follows from the same equation. In the ocean the horizontal
flow is nearly non-divergent, so $\partial u/\partial\xi$ and $\partial
v/\partial\eta$ are typically much larger than their sum: the horizontal
divergence is a small residual of two nearly cancelling terms. A
finite-difference estimate of it, which depends on the grid's horizontal
spacing and metric terms, therefore carries a relative error that can be
large compared with the divergence itself, and diagnosing $w$ by
integrating this estimate in the vertical accumulates that error from
level to level. Diagnostic reconstructions of $w$ from
Equation~\ref{eq:continuity} alone can thus carry substantial numerical
uncertainty irrespective of the physical validity of the equation itself.
This is part of the practical motivation for predicting $w$ directly from
$(u,v)$ with a data-driven model, rather than relying solely on its
diagnostic evaluation from continuity.

\subsection{Vertical correlation of the vertical velocity}
\label{sec:coherence}

Equation~\ref{eq:continuity} is not only linear but also local: it relates the
horizontal divergence at a point to the vertical derivative $\partial
w/\partial z$ at the same point. Obtaining $w$ itself requires integrating
this relation in the vertical, which links $w$ at different depths, but
the integral involves the horizontal divergence at all the intermediate
depths, which a stencil taken at a single level does not provide. Whether a horizontal $(u,v)$ stencil measured at one
reference depth can be informative about $w$ several levels below depends
on how correlated the vertical velocity field actually is across depth in
the dataset used here. We characterize this directly, before fitting
any predictive model, by computing the vertical autocorrelation of $w$ on each
of the domains used in this manuscript.

For a pair of levels $(k_1,k_2)$, we compute the spatial Pearson correlation
between $w_{k_1}$ and $w_{k_2}$ over the horizontal grid, using only the
grid points at which $w$ is defined at both levels (the valid set
$\Omega_{k_1,k_2}$ is recomputed independently
for every pair), expressed in terms of the covariance and standard
deviations computed over that same set:
\begin{equation}
\begin{split}
  r(k_1,k_2) &= \frac{\mathrm{Cov}(w_{k_1},w_{k_2})}{\sigma_{k_1}\,\sigma_{k_2}} , \\
  \mathrm{Cov}(w_{k_1},w_{k_2}) &= \frac{1}{|\Omega_{k_1,k_2}|}
     \sum_{p\in\Omega_{k_1,k_2}} \delta w_{k_1,p}\,\delta w_{k_2,p} ,
\end{split}
  \label{eq:pearson}
\end{equation}
with $\delta w_{k,p} \equiv w_{k,p}-\bar w_{k}$, where
$w_{k,p}$ is $w$ at level $k$ and horizontal grid point $p$, $\bar w_{k_1},\bar
w_{k_2}$ are the means taken over $\Omega_{k_1,k_2}$ only, and
$\sigma_{k} = [\mathrm{Cov}(w_{k},w_{k})]^{1/2}$. From the full $(k_1,k_2)$
matrix we also extract the profile of the correlation of $w$ at each
level with $w$ at a fixed near-surface reference level, as a function of
depth. This profile measures how far below the reference level $w$
remains linearly correlated with its value there, and therefore
bounds the expected reach of a stencil taken at a single near-surface
level.

\subsection{Validation domains}
\label{sec:domains}

Figure~\ref{fig:zlevels_domains}b shows the three domains used in this manuscript,
overlaid on the CROCO bathymetry: the South-West Mediterranean training
domain, the Alboran validation domain, and the transition validation domain.

The models are trained on the South-West Mediterranean domain
($\mathrm{lon}\in[0.40,2.00]^\circ$, $\mathrm{lat}\in[36.90,38.30]^\circ$)
and validated out of sample on two independent domains, geographically
separated from the training domain and from each other: the Alboran Sea
($\mathrm{lon}\in[-3.60,-2.20]^\circ$, $\mathrm{lat}\in[35.70,36.40]^\circ$)
and a transition region in the open sea between Almer\'{\i}a and
Cartagena ($\mathrm{lon}\in[-1.50,-0.20]^\circ$,
$\mathrm{lat}\in[36.40,37.40]^\circ$), referred to below as the
transition domain.
All three are cut from the same CROCO grid and restricted to ocean grid
points. Two considerations motivate using two
validation domains rather than one. First, Alboran is also used during
training, as an early check of the out-of-sample performance that decides when
training stops; it is therefore not a fully independent test, and the
performance reported on it may be optimistically biased. The transition
domain is not used during training and provides a fully independent test. Second, the
Alboran Sea is dynamically peculiar: it is the outflow region of the
Atlantic jet through the Strait of Gibraltar, with energetic gyres,
fronts and filaments, and its $w$ field also contains contributions from
tides and topographic internal waves \cite{Capo2021}, which are not set by
the local near-surface $(u,v)$. Validating there alone may therefore
understate the performance of the models elsewhere. The transition domain
lies between the Alboran Sea and the open western Mediterranean, between
the Almer\'{\i}a--Oran and the Cartagena--T\'en\`es
fronts \cite{HernandezCarrasco2018}. Its bathymetry is close to that of
the training domain, a deep basin with a median depth of $2633$~m, against
$2744$~m for the training domain (and $1018$~m for Alboran), and it does not share the peculiarities
of Alboran; it still differs from the training domain because of the
strong influence of the incoming Atlantic water. It can also be reached by
the energetic mesoscale eddies generated by instability of the Algerian
Current \cite{Millot1990,dOvidio2009}.

Validating on whole, spatially separate domains, rather than on grid
points withheld at random from the training domain, is required by the
models used here and not only a stricter test. A random
split is common practice in machine learning when the samples are
independent of one another, which is not the case here: neighboring grid
points are strongly correlated, and the models predict $w$ at a grid point
from the horizontal velocity at that point and at its nearest neighbors.
The training field must therefore be complete and contiguous: withholding
points at random would leave gaps in the neighborhoods of the points
around them. Along the same lines,
\citeA{HeMahadevan2024} reserve an entire snapshot of each of
their simulations for testing, rather than individual grid points.

The performance of the models is evaluated, on each domain and separately
at each target level, with the coefficient
\begin{equation}
  R^2 = 1-\frac{\sum_{n=1}^{N}\left(w_n-\hat w_n\right)^2}
               {\sum_{n=1}^{N}\left(w_n-\bar w\right)^2},
  \label{eq:r2}
\end{equation}
where the sums run over the $N$ valid grid points of the domain at that
level, $w_n$ is the CROCO vertical velocity, $\hat w_n$ its predicted value,
and $\bar w$ the mean of $w_n$ over the same points. $R^2$ is a standard
metric for evaluating regression models \cite{Chicco2021}: it equals $1$
for a perfect prediction and $0$ for one no better than the domain mean,
and, despite what the notation $R^2$ suggests, it becomes negative when
the prediction is worse than the domain mean.

\subsection{Nearest-neighbor stencil and per-level linear regression}
\label{sec:lr}

As discussed in Section~\ref{sec:continuity}, diagnosing $w$ by integrating
a finite-difference estimate of the horizontal divergence in the vertical
accumulates its discretization and grid-metric errors. We therefore do not
evaluate the divergence explicitly: the regression learns the mapping
from the raw $(u,v)$ stencil to $w$ directly from the data.

The linear baseline predicts $w$ at a grid point $(j,i)$, where $j$ and
$i$ are the grid indices along $\eta$ and $\xi$ (approximately
south--north and west--east), from the horizontal velocity field in a
$(2r+1)\times(2r+1)$ nearest-neighbor stencil centered on that point: with $\mathcal N_r=\{-r,-r+1,\dots,r\}^2$, the stencil collects
$u_{j+\Delta j,\,i+\Delta i}$ and $v_{j+\Delta j,\,i+\Delta i}$ for every
$(\Delta j,\Delta i)\in\mathcal N_r$ at a fixed reference level
$k_{\rm start}$ close to the surface, so that a stencil of radius $r$ spans
$(2r+1)^2$ grid points and $2(2r+1)^2$ features.
A stencil is valid, and its central grid point is used, only when all of
its $(2r+1)^2$ grid points lie in the water, i.e.\ none of them falls on
land or outside the domain. This requirement follows from what the
models use: the horizontal structure of the $(u,v)$ field around a
point, not the value of the field at the point itself. A grid point at
which $w$ is defined but part of the surrounding stencil is missing is
therefore discarded, rather than kept with an incomplete neighborhood.
With this definition the training domain contains $83792$ valid
stencils, the Alboran domain $36379$ and the transition domain
$48333$.

If $w$ is correlated across a range of nearby depths
(Section~\ref{sec:coherence}),
this single, near-surface $(u,v)$ stencil should already carry information
predictive of $w$ several levels below the surface, without needing
deeper horizontal-velocity measurements. We test this directly: the
same stencil at $(j,i)$, built exclusively from level $k_{\rm start}$, is
used as input for every target level, but an independent
ordinary-least-squares fit is obtained separately for each of the $N_{\rm target}=16$ target
levels $k$, from the stencil level $k_{\rm start}$ downwards,
\begin{equation}
\begin{split}
  w_{k}(j,i) = \sum_{(\Delta j,\Delta i)\in\mathcal N_r}
     \Big( a^{(k)}_{\Delta j,\Delta i}\,u_{j+\Delta j,\,i+\Delta i} \\
         + b^{(k)}_{\Delta j,\Delta i}\,v_{j+\Delta j,\,i+\Delta i}\Big)
     + c^{(k)} ,
\end{split}
  \label{eq:stencil_model}
\end{equation}
where $a^{(k)}$ and $b^{(k)}$ are the coefficients of $u$ and $v$ and
$c^{(k)}$ is the intercept, each fit using as target only the values of
$w$ at its own level. This is the
simplest possible model, and it serves two purposes: as a
physically-interpretable baseline in its own right, and as a reference
against which the FiLM network introduced below must be
judged. Its performance as a function of depth also shows how far below
the stencil level the near-surface $(u,v)$ remains linearly informative about $w$.

\subsection{Depth as a network input: a depth-conditional CNN}
\label{sec:cnn}

The per-level linear model of Section~\ref{sec:lr} is, by construction, not a
single continuous function of depth: it requires an independent fit
$\{a^{(k)},b^{(k)},c^{(k)}\}$ for every target level $k$, and cannot
be queried at a depth it was not explicitly fit to. Moreover, the fitted
coefficients are in general different at every depth, so a single model spanning
the whole target column must, in general, depend on depth $z$ in a way
that is not itself linear, even though the continuity equation that
motivates the stencil is linear at any fixed depth. Generalizing the
predictor to an arbitrary target depth $z$ therefore requires
supplying $z$ to the model as an explicit, free input parameter, and turns
the estimation problem into a genuinely non-linear regression task, which
motivates moving from the linear baseline to a neural network.

The stencil representation of Equation~\ref{eq:stencil_model} is naturally
reinterpreted as a two-channel image, $(u,v)$ over a $(2r+1)\times(2r+1)$
window, which a CNN can process directly
rather than through a flattened linear map.

Concretely, each training example pairs two inputs, the $7\times7$
two-channel $(u,v)$ stencil taken at the fixed reference level
$k_{\rm start}=2$ ($z_{\rm start}=-11.2$~m) around a horizontal grid point and a
target depth $z$, one of the $N_{\rm target}=16$ levels between $-11.2$
and $-116.6$~m, with the model value of $w$ at the center of the stencil
at that depth, which serves only as the regression target in the training
loss. Since every stencil is paired with all $16$ target depths, the
$83792$ valid training stencils (i.e.\ with no land point in the
stencil) yield $1340672$ training rows, and
any depth dependence of the network output can enter only through $z$.

All inputs and targets are standardized with statistics of the training
domain: $u$ and $v$ with their mean and standard deviation over all
training stencils, $z$ with the mean and standard deviation of the target
depths, and $w$ separately at each level, with the mean and standard
deviation of $w$ at that level, so that all levels contribute comparably
to the loss. Predictions are converted back to m\,s$^{-1}$ with the same
per-level statistics.

\begin{figure*}[!htbp]
  \centering
  \includegraphics{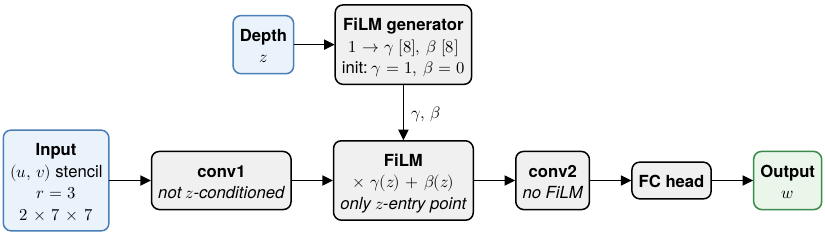}
  \caption{Architecture of the FiLM-conditioned CNN. The input is the
    $(u,v)$ stencil of radius $r=3$ ($7\times7$ grid points at the
    reference level $k_{\rm start}$, Section~\ref{sec:lr}), i.e.\ two
    channels of $7\times7$ values. The first convolutional layer,
    \texttt{conv1}, is a Conv2d layer from 2 to 8 channels with a
    $3\times3$ kernel and padding 1, so that its output keeps the
    $7\times7$ size, followed by ReLU and by Dropout2d with probability
    $0.1$. The FiLM layer multiplies each of the 8 output channels of
    \texttt{conv1} by $\gamma(z)$ and adds $\beta(z)$
    (Equation~\ref{eq:film}). The second convolutional layer, \texttt{conv2},
    is a Conv2d layer from 8 to 16 channels with a $3\times3$ kernel and
    no padding, which reduces the size to $5\times5$, followed by ReLU and
    Dropout2d with probability $0.1$; it is not modulated by FiLM. The
    fully-connected head flattens the $16\times5\times5$ output of
    \texttt{conv2} into a vector of 400 values and applies a Linear layer
    from 400 to 16 values, ReLU, Dropout with probability $0.2$ and a
    Linear layer from 16 values to the predicted $w$. The depth $z$ enters
    the network only through the FiLM generator, a Linear layer from 1 to
    16 values that gives $\gamma$ and $\beta$ for the 8 channels; neither
    convolutional layer depends on $z$. The FiLM generator is initialized
    to the identity ($\gamma=1$, $\beta=0$) and regularized with a weight
    decay of $5\times10^{-5}$, 20 times weaker than the weight decay of
    $10^{-3}$ applied to the rest of the network. The network has 7785
    parameters, most of them in the first fully-connected layer. Layer
    names used in this caption are those of the PyTorch library \protect\cite{Paszke2019}, in which the
    network is implemented: Conv2d is its two-dimensional convolutional
    layer (\texttt{torch.nn.Conv2d}), Linear its fully-connected layer
    (\texttt{torch.nn.Linear}), and ReLU, Dropout2d and Dropout the
    corresponding \texttt{torch.nn} modules.}
  \label{fig:film_architecture}
\end{figure*}

To predict $w$ at an arbitrary target depth $z$ with a single network,
depth is supplied to the network through Feature-wise Linear Modulation
(FiLM) \cite{Perez2018}. The FiLM network comprises two convolutional
layers, with a single FiLM modulation between them, and a small
fully-connected head (Figure~\ref{fig:film_architecture}). The first
convolutional layer (\texttt{conv1}, using `same' padding) transforms the
two input fields, $u$ and $v$ on the $7\times7$ stencil, into $C=8$ new
fields on the same $7\times7$ grid, each obtained by combining, with
learned weights, the values of $u$ and $v$ in a $3\times3$ window of
neighboring grid points (the convolution kernel); in machine-learning
terminology these fields are called feature maps, or channels. This layer
receives no FiLM modulation, so that its output is the same for all
target depths. Denoting by $\bm{x}_c$ the $c$-th feature map, an array of
$7\times7$ values, FiLM then rescales and shifts each map by an amount
that depends on the target depth,
\begin{equation}
\begin{split}
  {\rm FiLM}(\bm{x})_c &= \gamma_c(z)\,\bm{x}_c + \beta_c(z) ,
  \qquad c=1,\dots,C , \\
  \begin{pmatrix}\bm{\gamma}(z)\\ \bm{\beta}(z)\end{pmatrix} &= \bm m_{\rm FiLM}\, z + \bm q_{\rm FiLM} ,
\end{split}
  \label{eq:film}
\end{equation}
where $\gamma_c(z)$ and $\beta_c(z)$ are a scale factor and an offset for
map $c$, the same at all the $7\times7$ grid points, collected in the
vectors $\bm\gamma(z),\bm\beta(z)\in\mathbb{R}^{C}$. These $2C$
coefficients are linear functions of the scalar depth $z$, with slopes
$\bm m_{\rm FiLM}\in\mathbb{R}^{2C}$ and intercepts
$\bm q_{\rm FiLM}\in\mathbb{R}^{2C}$. Like every other weight of the network,
$\bm m_{\rm FiLM}$ and $\bm q_{\rm FiLM}$ are free parameters, learned
together with the convolutional layers by minimizing the same prediction
error, rather than fixed or set by hand. They are initialized so that
$\gamma_c(z)=1$ and $\beta_c(z)=0$ at the start of training, i.e.\ FiLM
initially leaves the maps unchanged. This modulation is the only point at
which depth enters the model. The second convolutional layer
(\texttt{conv2}, unpadded) combines the $8$ depth-modulated maps, again in
$3\times3$ windows, into $16$ new maps on a $5\times5$ grid, with weights
that do not depend on $z$, and the fully-connected head maps the result
to $w$. Because every layer downstream of the FiLM modulation depends on
$z$, each (stencil, level) pair still enters the loss as a separate term,
with its own target and its own modulation. The first convolutional
layer, which does not see $z$, is instead evaluated only once per stencil
and its output shared by the $N_{\rm target}=16$ levels of that stencil,
and mini-batches are built accordingly from whole stencils: $128$
stencils with all their $16$ levels, i.e.\ $2048$ (stencil, level) pairs,
per batch.

This two-layer configuration was reached by reducing an initial,
three-layer prototype (channel widths $32/64/64$, FiLM applied after
every convolutional layer, $269921$ parameters) that showed
clear signs of overfitting; the
current network has $7785$ parameters, most of them in the
first fully-connected layer. Following this pattern, three further
regularization mechanisms are used during training: dropout \cite{Srivastava2014}, applied
per-channel ($p=0.1$) on the convolutional feature
maps and per-unit ($p=0.2$) on the fully-connected head; an $L_2$
weight-decay penalty \cite{Krogh1991}, $10^{-3}$ for the rest of the network but only
$5\times10^{-5}$ for the FiLM generator's own parameters
$\bm m_{\rm FiLM},\bm q_{\rm FiLM}$; and the stopping rule described below. The penalty on the FiLM generator is weaker because an $L_2$
penalty pulls $\bm m_{\rm FiLM}$ and $\bm q_{\rm FiLM}$ toward zero, i.e.\
$\gamma$ toward $0$ rather than toward the identity, so a strong one
would shrink the modulated feature maps as a whole. The FiLM network is trained with Adam \cite{Kingma2015} on the mean-squared
error of the standardized target, starting from a learning rate of
$10^{-3}$ that is halved whenever the training loss plateaus for $10$
epochs. Training is stopped when the learning rate reaches its minimum of
$10^{-6}$, chosen a priori, or when the mean
out-of-sample $R^2$ on the Alboran validation domain, evaluated at every
epoch up to the twenty-fifth and every $25$ epochs thereafter, has not
improved for $300$ epochs. Alboran data are thus used only to decide when training stops:
the weights that are evaluated are those of the last epoch, not those that
score best on Alboran.
The out-of-sample performance of a single trained network depends on the random
seed, which sets the weight initialization, the order of the training
batches and the dropout masks. To quantify this numerical uncertainty, the
FiLM network was trained $274$ times, with identical data, architecture and
hyperparameters and different seeds; all FiLM results below are reported
as the mean and sample standard deviation $\sigma$ of $R^2$ over this
ensemble. The standard deviation $\sigma$ is the numerical uncertainty of
a single trained network. The uncertainty of the ensemble mean is the
standard error, i.e.\ $\sigma$ divided by the square root of the number
of realizations, which is at most $0.01$ for all the
values reported here. Note that the per-level linear regression of Section~\ref{sec:lr}
is a closed-form least-squares fit and is therefore deterministic.

The FiLM network is trained on the South-West Mediterranean
domain. It is then applied to the Alboran and
transition domains with the weights obtained in training, unchanged, as is
the linear baseline of Section~\ref{sec:lr}.

\section{Results}
\label{sec:results}

\subsection{Incompressibility check}

As a first check, we evaluate the three-dimensional divergence
$\nabla_{\xi\eta z}\cdot\mathbf{u}$ of the CROCO fields at every valid
grid point of the South-West Mediterranean training domain. Spatial
derivatives are computed on the interpolated fields with a
second-order centered finite-difference scheme (first-order one-sided
differences next to land), along the grid directions $(\xi,\eta)$ in the
horizontal and across the fixed $z$-levels in the vertical. In the
simulation, which is hydrostatic, $w$ is not a prognostic variable: at
every time step it is diagnosed from the discrete continuity equation on
the native $\sigma$-grid, by summing, grid cell by grid cell, the
divergence of the horizontal fluxes upward from the sea floor, where the
vertical flux vanishes \cite{ShchepetkinMcWilliams2005}. On that
grid the model $w$ therefore satisfies the incompressibility equation
exactly, and any departure of $\nabla_{\xi\eta z}\cdot\mathbf{u}$ from
zero is introduced by our post-processing, i.e.\ by the interpolation
onto the fixed levels and by the numerical derivatives.

\begin{figure}[!htbp]
  \centering
  \includegraphics{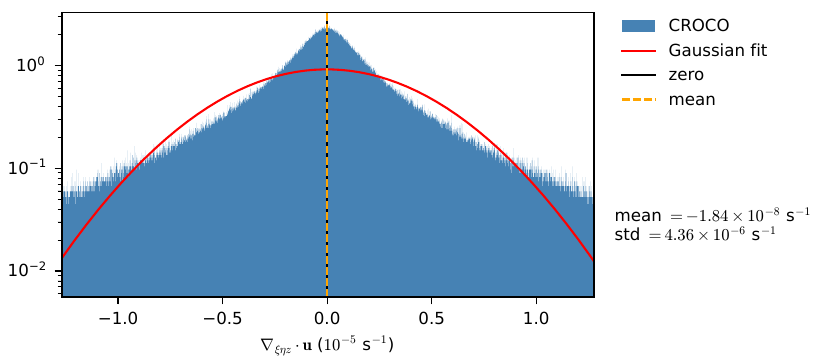}
  \caption{Probability density function (log scale, in units of
    $10^{5}$~s) of
    $\nabla_{\xi\eta z}\cdot\mathbf u$ over all valid grid
    points of the CROCO South-West Mediterranean domain, with a Gaussian
    fit overlaid. Mean $-1.84\times10^{-8}\,{\rm s}^{-1}$, standard
    deviation $4.36\times10^{-6}\,{\rm s}^{-1}$.}
  \label{fig:divergence}
\end{figure}

Figure~\ref{fig:divergence} shows the distribution of
$\nabla_{\xi\eta z}\cdot\mathbf u$. It is centered very close to zero, with mean
$-1.84\times10^{-8}\,{\rm s}^{-1}$, but its standard deviation,
$4.36\times10^{-6}\,{\rm s}^{-1}$, is more than two orders of magnitude
larger. The incompressibility equation is therefore satisfied on
average, but not point by point. In our case the derivative field carries
the error of the interpolation onto the fixed levels, which removes the
exactness of the discrete balance of the model, and the residual of the
order of $10^{-6}\,{\rm s}^{-1}$ at each grid point is its signature.
Moreover, the horizontal divergence is a small residual of much larger
derivative terms (Section~\ref{sec:continuity}), so that small errors in
each of them produce a large relative error in their sum. Both effects
would propagate into $w$ if it were diagnosed by integrating the
horizontal divergence in the vertical, and they justify not using the
derivative field: the models of Sections~\ref{sec:lr} and~\ref{sec:cnn}
take the raw $(u,v)$ stencil as input.

\subsection{Vertical correlation of $w$}
\label{sec:wcoherence}

\begin{figure*}[!htbp]
  \centering
  \includegraphics{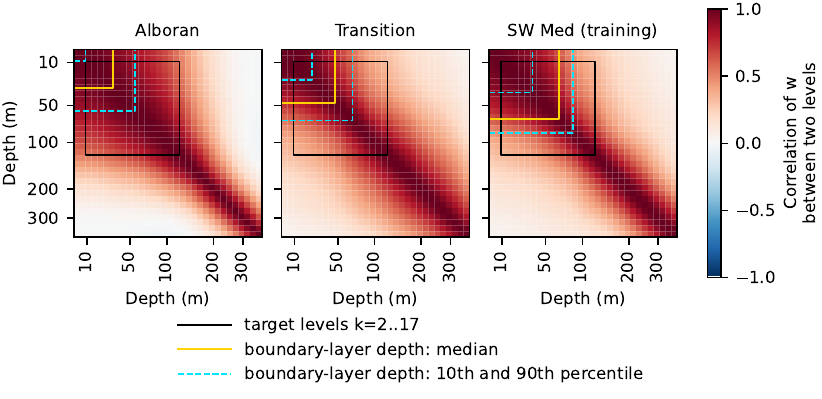}
  \caption{Level-by-level Pearson correlation of $w$ on the Alboran and
    transition validation domains and on the South-West Mediterranean
    training domain, on fixed $z$-levels over the upper $400$~m. Each cell
    is one model level: the cells are equally spaced along the axes, one
    per level, and the axis ticks give the corresponding depth, so the
    depth scale is not linear. The black square marks the target levels
    $k=2$--$17$ ($-11.2$ to $-116.6$~m). The yellow squares mark the
    median and the dashed cyan squares the $10$th and $90$th percentiles,
    over the grid points of each domain, of the depth $h_{\rm bl}$ of the
    surface boundary layer diagnosed by the model \protect\cite{Capo2021}.}
  \label{fig:coherence}
\end{figure*}

Figure~\ref{fig:coherence} shows the level-by-level correlation matrix of
$w$ on the three domains, computed on fixed $z$-levels over the upper
$400$~m. In all three a band of high correlation lies along the diagonal:
successive levels are almost perfectly correlated, with $r\geq0.96$
between any two adjacent levels within the target window, and the
correlation decays as the two levels move apart.

\begin{figure}[!htbp]
  \centering
  \includegraphics{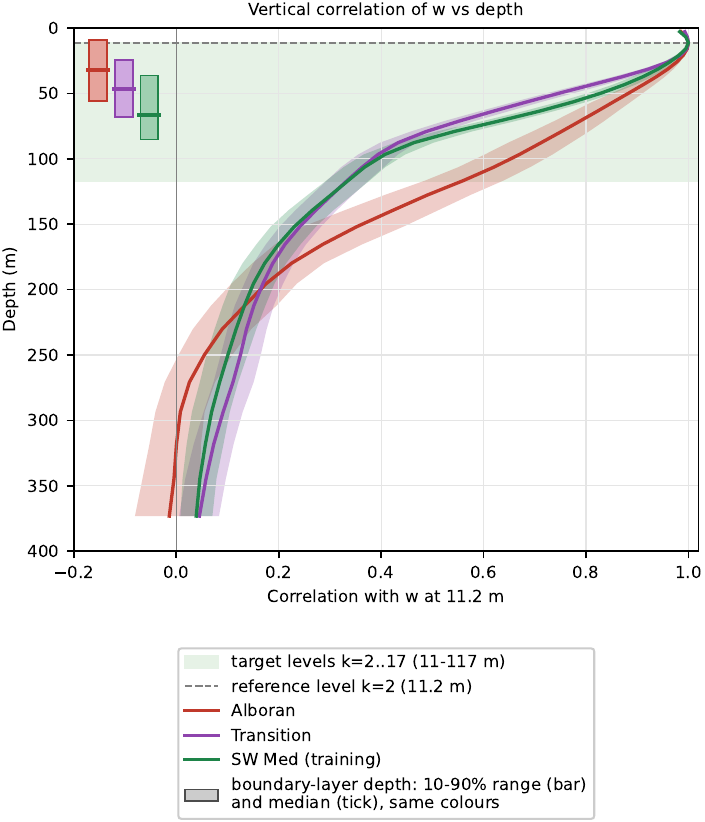}
  \caption{Spatial correlation of $w$ at each depth with $w$ at the
    stencil level $k_{\rm start}=2$ ($z_{\rm start}=-11.2$~m, dashed line), for the
    three domains. Green shading: target levels $k=2$--$17$. Vertical bars
    on the left: depth $h_{\rm bl}$ of the surface boundary layer diagnosed
    by the model \protect\cite{Capo2021} on each domain (same colors as the
    curves); the bar spans the $10$--$90\%$ range over the grid points of
    the domain and the tick marks the median. Shaded band around each
    curve: $95\%$ confidence interval of the correlation, from a block
    bootstrap that resamples $10\times10$~km blocks of grid points, to
    account for the spatial correlation of $w$.}
  \label{fig:corrdepth}
\end{figure}

The quantity of direct interest here
is the row of the matrix at the stencil level, i.e.\ the correlation of
$w$ at each depth with $w$ at $k_{\rm start}=2$ ($z_{\rm start}=-11.2$~m), shown in
Figure~\ref{fig:corrdepth}. Within the target levels used in this manuscript
($N_{\rm target}=16$ levels between $-11.2$ and $-116.6$~m,
Section~\ref{sec:cnn}) the loss of correlation is already substantial: at the
deepest target level the correlation is $0.56$ on Alboran and
$0.33$--$0.34$ on the other two domains, with $95\%$ confidence intervals
of $0.49$--$0.64$ on Alboran and $0.29$--$0.38$ on the other two
(Figure~\ref{fig:corrdepth}). The depth at which it drops to
one half, interpolated between levels, is about $125$~m on Alboran,
against about $78$ and $84$~m on the transition and training domains. Below the target window the
correlation keeps decreasing, and at $-294$~m it is below $0.1$ on all
three domains. This decay foreshadows the degradation of predictive performance
with depth reported below.

The vertical structure of the correlation can be compared with the
surface boundary layer,
i.e.\ the depth down to which surface-forced turbulence is active. Its
depth $h_{\rm bl}$ is diagnosed by the model at every time step through
its K-profile parameterization of vertical mixing \cite{Large1994}, and
is the quantity shown by \citeA{Capo2021} for the same
simulation. In the snapshot used here the median
$h_{\rm bl}$ is $32$~m on Alboran ($10$--$90\%$
range $9$--$56$~m; Figures~\ref{fig:coherence} and~\ref{fig:corrdepth}), $47$~m on the transition domain ($25$--$68$~m) and $66$~m on the
South-West Mediterranean training domain ($36$--$85$~m). The stencil level
($-11.2$~m) therefore lies inside the boundary layer almost everywhere
($87$--$98\%$ of the grid points, depending on the domain), whereas the
target levels progressively leave it: $k=7$ ($-37.1$~m) lies inside it at
$39$--$89\%$ of the points, $k=12$ ($-71.1$~m) at only $3$--$38\%$, and
$k=17$ ($-116.6$~m) nowhere. On the transition and training domains the
depth at which the correlation drops to one half lies close to the base
of the deepest boundary layers (the $90$th percentile of $h_{\rm bl}$, $68$
and $85$~m), whereas on Alboran it is about twice as deep as that ($56$~m). A similar depth dependence was found by
\citeA{HeMahadevan2024} in their simulations, where the performance of
their models was tested in relation to the mixed layer depth. They found
this performance to be high above the base of the mixed layer and to drop
towards zero below it.

A plausible physical origin for the high correlation of $w$ within the
boundary layer is its strong surface-forced turbulence, parameterized in
the model as vertical mixing, which exchanges momentum between the fluid
levels. Vertical motions that span the upper water
column can extend the correlation below the boundary layer. At the
submesoscales resolved by the simulation, one candidate is the
ageostrophic secondary circulation of fronts and filaments, driven by
frontogenesis and by the turbulent thermal-wind
balance \cite{McWilliams2016,Capo2021}; others are tides and internal
waves, and, at larger scales, the Ekman pumping induced by the
interaction of mesoscale eddies with the wind
stress \cite{Stern1965,Gaube2015}, and the vertical motions associated
with the formation, intensification and decay of mesoscale
eddies \cite{McGillicuddy2016}. This is consistent with what is seen on Alboran, where frontal
downwelling, tides and internal waves are all prominent \cite{Capo2021}
and where, as shown above, the correlation with the stencil level extends
about twice as deep as the boundary layer, unlike on the other two
domains.

\subsection{Linear regression baseline and coefficient interpretability}

\begin{figure*}[!htbp]
  \centering
  \includegraphics{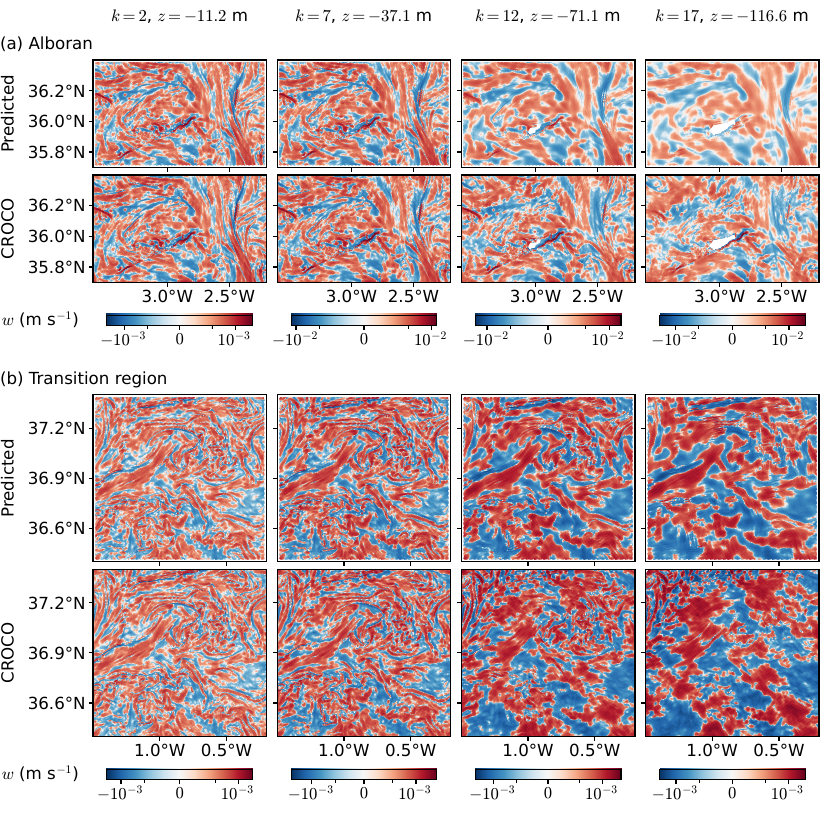}
  \caption{Per-level linear regression of Section~\ref{sec:lr}: predicted
    (upper row of each panel) vs.\ CROCO (lower row) horizontal maps of $w$ on the two
    independent validation domains, (a) Alboran and (b) the transition region, at the
    four target levels compared in Table~\ref{tab:lr_vs_film} (columns, $k=2,7,12,17$; $z=-11.2$ to
    $-116.6$~m). The model was fitted exclusively on the South-West
    Mediterranean training domain and applied here with the same
    coefficients. The white patch in the Alboran panels is Albor\'an
    Island.
    The color scales are symmetric logarithmic.}
  \label{fig:lr_val_maps}
\end{figure*}

Table~\ref{tab:lr} collects the $R^2$ of the per-level independent linear
regression (stencil radius $r=3$, i.e.\ $7\times7$, $98$ features;
$83792$ valid South-West Mediterranean training grid points, i.e.\ with no
land point in the stencil), in-sample and out-of-sample on the
independent Alboran and transition validation domains, at every target
level. Consistent with
the correlation analysis above, in-sample $R^2$ is highest near the
reference level ($R^2=0.960$ at $z=-20.8$~m) and
degrades monotonically below it, falling to $R^2=0.409$ at
the deepest level of the target window, $z=-116.6$~m. Out of sample the
profile has the same shape and peaks at the same level ($0.963$ on
Alboran and $0.962$ on the transition domain), with values comparable to the in-sample
ones over the upper levels, but it degrades faster with depth: at
$z=-71.1$~m $R^2$ is $0.546$ on Alboran and $0.398$ on the transition domain, against
$0.628$ in sample, and at $z=-116.6$~m it is $0.183$ and $0.201$, against
$0.409$. Figure~\ref{fig:lr_val_maps} compares the predicted and CROCO
$w$ on the two validation domains at four of these levels.

\begin{table}[!htbp]
\centering
\begin{tabular}{ccccc}
\toprule
 & & In-sample & \multicolumn{2}{c}{Out-of-sample} \\
\cmidrule(lr){3-3}\cmidrule(lr){4-5}
$k$ & $z$ (m) & SW Med. & Alboran & Transition \\
\midrule
2  & $-11.2$    & 0.898 & 0.924 & 0.922 \\
3  & $-15.9$    & 0.942 & 0.952 & 0.948 \\
4  & $-20.8$    & 0.960 & 0.963 & 0.962 \\
5  & $-25.9$    & 0.956 & 0.954 & 0.953 \\
6  & $-31.4$    & 0.938 & 0.929 & 0.916 \\
7  & $-37.1$    & 0.908 & 0.890 & 0.849 \\
8  & $-43.2$    & 0.868 & 0.840 & 0.758 \\
9  & $-49.6$    & 0.818 & 0.782 & 0.656 \\
10 & $-56.3$    & 0.759 & 0.714 & 0.556 \\
11 & $-63.5$    & 0.694 & 0.636 & 0.467 \\
12 & $-71.1$    & 0.628 & 0.546 & 0.398 \\
13 & $-79.1$    & 0.568 & 0.448 & 0.343 \\
14 & $-87.6$    & 0.522 & 0.355 & 0.296 \\
15 & $-96.6$    & 0.486 & 0.279 & 0.256 \\
16 & $-106.3$   & 0.453 & 0.222 & 0.226 \\
17 & $-116.6$   & 0.409 & 0.183 & 0.201 \\
\bottomrule
\end{tabular}
\vspace{10pt}
\caption{$R^2$ of the per-level independent linear regression ($r=3$, i.e.\
  a $7\times7$ stencil with $98$ features; $k_{\rm start}=2$,
  $N_{\rm target}=16$) at every target level: in-sample on the South-West
  Mediterranean training domain ($83792$ valid grid points, i.e.\ with no land point in the stencil) and out-of-sample on the two
  independent validation domains of Section~\ref{sec:domains} (Alboran,
  $N=35945$--$36379$ depending on level; transition domain,
  $N=48333$). See Figure~\ref{fig:lr_coefficients} for the corresponding
  fitted coefficient maps at $k=4$ and $k=10$.}
\label{tab:lr}
\end{table}

\begin{figure}[!htbp]
  \centering
  \includegraphics{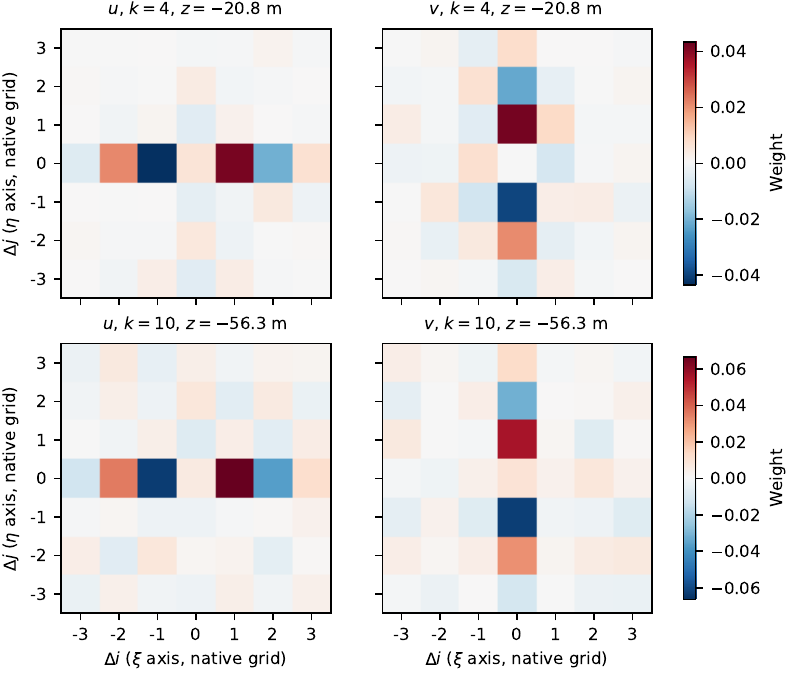}
  \caption{Fitted $u$- and $v$-coefficient maps of the per-level linear
    regression at $k=4$ ($z=-20.8$~m) and $k=10$
    ($z=-56.3$~m), in native grid coordinates $(\xi,\eta)$. The
    small-large-zero-large-small sign pattern along the active row/column
    is qualitatively consistent with, though not a quantitatively exact
    match to, the fourth-order centered finite-difference stencil of
    Equation~\ref{eq:4thorder}.}
  \label{fig:lr_coefficients}
\end{figure}

What justifies treating this simple model as a physically meaningful
baseline, rather than an arbitrary one, is the structure of the fitted
coefficients $a^{(k)},b^{(k)}$ themselves. Figure~\ref{fig:lr_coefficients}
shows the fitted $u$- and $v$-coefficient maps at $k=4$ and $k=10$. Both levels display the same qualitative
structure: along the central row of the $u$-coefficient map
($\Delta j = 0$), the fitted weights are small and positive at
$\Delta i=-2$, large and negative at $\Delta i=-1$, near zero at
$\Delta i=0$, large and positive at $\Delta i=+1$, and small and negative
at $\Delta i=+2$, and the analogous pattern holds for the $v$-coefficients
along $\Delta j$ at $\Delta i = 0$. This is not the signature of a simple
two-point centered difference (which would only populate
$\Delta i = \pm1$); it displays the qualitative sign structure of a
centered finite-difference stencil, broadly consistent with, though not a
quantitatively exact match to, the coefficients of the fourth-order
centered finite-difference stencil
\begin{equation}
  \frac{\partial u}{\partial\xi}\Big|_{\Delta j=0} \approx
     \frac{-u_{+2}+8u_{+1}-8u_{-1}+u_{-2}}{12\,\Delta\xi} ,
  \label{eq:4thorder}
\end{equation}
with the corresponding relation for $\partial v/\partial\eta$. A direct
check of the fitted coefficients against this stencil confirms the
antisymmetry ($a_{-2}\approx-a_{+2}$ and $a_{-1}\approx-a_{+1}$, to within
a few percent of the coefficient magnitude, at both $k=4$ and $k=10$)
expected of a derivative-like operator. In other words, without being
given any explicit derivative operator, the
regression re-discovers the qualitative, centered and antisymmetric,
structure of the horizontal divergence $\partial u/\partial\xi+\partial
v/\partial\eta$ entering the continuity equation (Equation~\ref{eq:continuity})
directly from the data. This confirms that the linear model is
not merely curve-fitting: it is recovering, from data alone, an operator
with the same qualitative structure as the discretized divergence entering
the continuity equation, which is what justifies treating it as a
physically-motivated baseline against which the FiLM network below is
compared.

\subsection{Depth-conditional CNN (FiLM)}

An initial, three-layer prototype of the FiLM network
($269921$ parameters, FiLM applied after every convolutional
layer) overfitted the training data: its in-sample $R^2$ was much higher
than its out-of-sample $R^2$, the textbook signature of a
model memorizing training-domain specifics rather than learning the
underlying $(u,v)\to w$ relationship. A central point in diagnosing this is that the raw number of training rows
overstates the available information. Each row is a (stencil, target
level) pair: it adds a new target value, but not a new input, since the
same stencil recurs, unchanged, across every target level. The distinct
horizontal configurations the network sees are only the unique stencils,
$83792$ for the training domain of Table~\ref{tab:lr}, and the targets attached to each of them are themselves
strongly correlated in depth (adjacent-level $r\geq0.96$,
Section~\ref{sec:wcoherence}), so that the $16$ target values attached to a
stencil carry much less information than $16$ independent values. Viewed this way, a
network with several times as many parameters as unique stencils
had, despite its nominally large row count, ample capacity to memorize
features specific to the training domain, and therefore to overfit.

This motivated reducing the network to the two-layer, $7785$-parameter
architecture described in Section~\ref{sec:cnn}, a reduction by a
factor of about $35$, together with the dropout, weight-decay, and
stopping criteria already described. All results are evaluated on the
FiLM network as it is at the epoch at which training stops.

Both validation domains are evaluated at every validation check, although
only Alboran enters the stopping criterion (Section~\ref{sec:cnn}).

Training lasted between $262$ and $432$
epochs (mean $333$); $161$ of the $274$ realizations stopped because the
learning rate reached its floor and the remaining $113$ on the Alboran
criterion, with the same final training loss, $0.249\pm0.003$
(standardized MSE), in both groups. The resulting out-of-sample $R^2$ is
collected in Table~\ref{tab:lr_vs_film}. The model reproduces the
qualitative depth-degradation pattern already seen for the linear baseline
and anticipated by the correlation analysis of Section~\ref{sec:wcoherence}:
performance is highest near the reference level and decreases with depth
offset, on both validation domains.

\begin{table*}[!htbp]
\centering
\begin{tabular}{cccccccc}
\toprule
 & & \multicolumn{2}{c}{In-sample} & \multicolumn{2}{c}{Alboran} & \multicolumn{2}{c}{Transition} \\
\cmidrule(lr){3-4}\cmidrule(lr){5-6}\cmidrule(lr){7-8}
$k$ & $z$ (m) & LR & FiLM & LR & FiLM & LR & FiLM \\
\midrule
2  & $-11.2$  & 0.898 & $0.878\pm0.012$ & 0.924 & $0.745\pm0.044$ & 0.922 & $0.871\pm0.012$ \\
7  & $-37.1$  & 0.908 & $0.906\pm0.011$ & 0.890 & $0.702\pm0.038$ & 0.849 & $0.840\pm0.009$ \\
12 & $-71.1$  & 0.628 & $0.724\pm0.008$ & 0.546 & $0.400\pm0.052$ & 0.398 & $0.490\pm0.016$ \\
17 & $-116.6$ & 0.409 & $0.574\pm0.009$ & 0.183 & $0.095\pm0.121$ & 0.201 & $0.241\pm0.025$ \\
\bottomrule
\end{tabular}
\vspace{10pt}
\caption{$R^2$ of the per-level linear regression of
  Section~\ref{sec:lr} (LR) and of the depth-conditional FiLM CNN of
  Section~\ref{sec:cnn} at the four target levels: in-sample on the
  South-West Mediterranean training domain (the LR values are those of
  Table~\ref{tab:lr}) and out-of-sample on both independent validation
  domains. FiLM values are
  mean $\pm$ standard deviation over $274$ realizations differing only in
  the random seed; the LR is deterministic. On Alboran the FiLM
  distribution is skewed towards low values, with medians $0.752$, $0.710$,
  $0.407$ and $0.130$. On Alboran at $k=17$ the difference between the two
  models is smaller than the spread of the FiLM network.}
\label{tab:lr_vs_film}
\end{table*}
 
\begin{figure}[!htbp]
  \centering
  \includegraphics{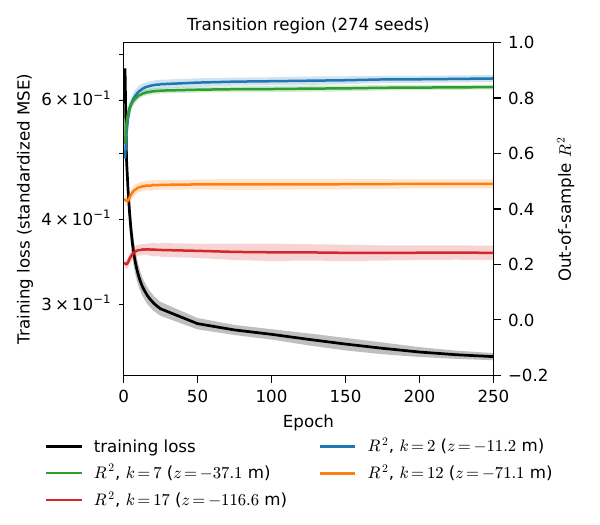}
  \caption{FiLM CNN, mean and standard deviation over the $274$
    realizations: training loss (standardized MSE, log scale, left axis)
    and out-of-sample $R^2$ on the transition domain at the four target levels (right
    axis), at every epoch up to the twenty-fifth and every $25$ epochs
    thereafter, up to epoch $250$, the last one reached by all
    realizations. Each line is the mean over the realizations, and the
    shaded band around it spans one standard deviation above and below
    the mean.}
  \label{fig:film_training}
\end{figure}

Figure~\ref{fig:film_training} shows how the training loss of the FiLM
network and its out-of-sample $R^2$ on the transition domain, the validation domain that
does not enter the stopping criterion, evolve over training, averaged over the
ensemble, at four target levels ($k=2,7,12,17$) spanning the training
column. Most of the optimization happens in the first few tens of epochs: the
training loss falls from $0.665\pm0.028$ at the first epoch to
$0.295\pm0.007$ by the twenty-fifth, and then only to $0.249\pm0.003$ at
the end of training, which for every realization comes after the last
epoch shown. The learning rate, initially $10^{-3}$, is halved every time the
training loss stops decreasing for $10$ epochs; the weight updates
therefore become smaller and smaller, and over the last part of the run
the weights remain nearly unchanged. After the first few tens of epochs
the out-of-sample $R^2$ on the transition domain becomes stationary at all four levels:
between the twenty-fifth epoch and epoch $250$ it changes by at most
$0.02$, from $0.850$ to $0.870$ at $k=2$, from $0.827$ to $0.839$ at
$k=7$, from $0.487$ to $0.490$ at $k=12$ and from $0.252$ to $0.241$ at
$k=17$. Most of the increase of $R^2$ takes place during the first
epochs, starting from $0.586$, $0.639$, $0.433$ and $0.203$ at the first
one. The ranking between levels is established by the fourth epoch and
does not change afterwards.

\begin{figure*}[!htbp]
  \centering
  \includegraphics{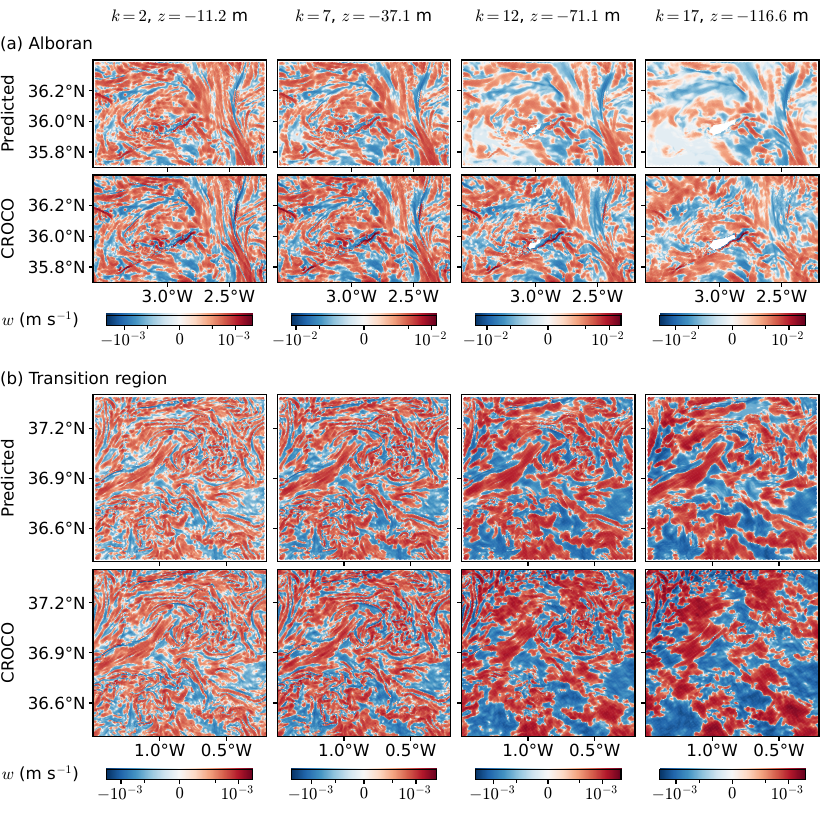}
  \caption{Depth-conditional FiLM CNN of Section~\ref{sec:cnn}: predicted
    (upper row of each panel) vs.\ CROCO (lower row) horizontal maps of $w$ on the two
    independent validation domains, (a) Alboran and (b) the transition
    region, at the same four target levels as in
    Figure~\ref{fig:lr_val_maps}. The predictions come from the single
    network trained on the South-West Mediterranean domain and applied to
    both validation domains with its weights unchanged. The color scales
    are symmetric logarithmic, and the white patch in the Alboran panels is
    Albor\'an Island.
    Maps are shown for one realization of the ensemble (seed $261$), the one
    whose out-of-sample $R^2$ is closest to the ensemble medians; the
    ensemble statistics are given in Table~\ref{tab:lr_vs_film}.}
  \label{fig:film_val_maps}
\end{figure*}

With this reduced, regularized FiLM network, the comparison with the per-level
linear baseline of Section~\ref{sec:lr} is made on both validation domains
and with the numerical uncertainty of the FiLM network in hand
(Table~\ref{tab:lr_vs_film}; the maps predicted by the FiLM network are shown
in Figure~\ref{fig:film_val_maps}). The picture is domain- and
depth-dependent rather than uniform. At the two shallower levels the
linear baseline leads on both domains: on Alboran in every realization,
with $R^2=0.924$ against $0.745\pm0.044$ at $k=2$ and $0.890$ against
$0.702\pm0.038$ at $k=7$; on the transition domain with closer values, $0.922$ against
$0.871\pm0.012$ at $k=2$, in every realization, and $0.849$ against
$0.840\pm0.009$ at $k=7$, in $86\%$ of them. At the two deeper levels the
two domains behave differently. On Alboran the linear baseline still
leads, in every realization at $k=12$ ($0.546$ against $0.400\pm0.052$)
and in $82\%$ of them at $k=17$ ($0.183$ against $0.095\pm0.121$); on
the transition domain the FiLM network is better, in every realization at $k=12$
($0.490\pm0.016$ against $0.398$) and in $93\%$ of them at $k=17$
($0.241\pm0.025$ against $0.201$). In other words, once the
FiLM network is regularized enough to generalize, its added
flexibility does not pay off near the reference level on either
validation domain, where the linear, physically-motivated operator of
Section~\ref{sec:lr} already captures essentially all of the achievable
performance; but at the two deepest levels tested, on the more dynamically
typical transition domain, the FiLM network continues to show
a genuine net gain over that linear baseline, a gain that is
absent on the atypical Alboran domain. Whether this pattern strengthens further
with a wider stencil or more training data is left for future work.

Taken together, the picture is the following. At the shallower levels,
where the stencil sits close to the target depth, the per-level linear
regression of Section~\ref{sec:lr} captures all the out-of-sample performance
available on both domains, and the additional flexibility of the FiLM
network brings no gain. These
are also the levels at which $w$ is still strongly correlated with its
value at the stencil level ($r\geq0.87$ down to $-37.1$~m on all three
domains, Section~\ref{sec:wcoherence}), whereas the drop in performance at $k=12$
and $k=17$ goes with the loss of that correlation. The
FiLM network does better only further down: at the two deepest levels
tested it overtakes the
linear baseline on the transition domain, while on Alboran it at best
matches it. Beyond the comparison of $R^2$ values, however, the FiLM network
provides something the per-level baseline cannot: a single model,
continuous in $z$, that can be queried at any depth within the training
range instead of being refitted level by level. As a check of that
continuity, the FiLM network is evaluated at $z=-52.96$~m, midway between the
training levels $k=9$ and $k=10$ and never itself used as one. Since the
CROCO $w$ is available only on the fixed levels, $R^2$ is computed against $w$
interpolated linearly between these two known levels, i.e.\ their average.
It is $0.562\pm0.043$ on Alboran and $0.664\pm0.011$ on the transition domain, in both
cases between the scores of the FiLM network at $k=7$ and $k=12$, the evaluated
levels that bracket this depth. This test does not validate the FiLM network
at a new depth, since the reference $w$ there is itself interpolated. It
shows, however, that between two training levels the performance of the
FiLM network does not drop, as it would if the network had learned only
the $16$ depths used in training: its predictions vary smoothly with $z$.

\section{Conclusions}
\label{sec:conclusions}

Motivated by the linear structure of the continuity equation, we have
introduced a linear regression baseline that maps a local
horizontal-velocity stencil onto $w$. The vertical velocity field $w$
decorrelates rapidly with depth: at the deepest target level
($-116.6$~m) its correlation with $w$ at the stencil level has fallen to
$0.56$ on Alboran and to $0.33$--$0.34$ on the other two domains. This
correlation is nevertheless enough for a per-level linear regression
built exclusively from a near-surface $(u,v)$ stencil to remain
predictive several levels below the surface. The fitted coefficients of this regression have the same
qualitative, centered and antisymmetric, structure as the discretized
horizontal divergence entering that same continuity equation, though not
its fourth-order finite-difference coefficients, which makes it a
physically interpretable and non-trivial baseline rather than an arbitrary
one. Because this linear model must be fit independently at
every target depth, generalizing it to an arbitrary depth requires
supplying depth itself as an explicit, free input to the model, which
turns the estimation problem into a genuinely non-linear one and motivates
a depth-conditional convolutional neural network, here conditioned on
depth through feature-wise linear modulation (FiLM). Overfitting proved
to be a first-order concern for this network when trained and validated
on a single domain: an initial, larger prototype reached a much higher
$R^2$ in sample than out of sample. Once this is addressed by substantially reducing
network capacity and adding standard regularization, the comparison with
the linear baseline is more nuanced than a simple win or loss for either
model (Section~\ref{sec:results}). Since a single
out-of-sample region can bias this comparison, we also validated the
models on a second, independent transition domain in the open sea between
Almer\'{\i}a and Cartagena ($36.4$--$37.4^\circ$N, $1.5$--$0.2^\circ$W;
Section~\ref{sec:domains}), less dynamically anomalous than the Alboran
outflow region. Near the reference depth, essentially all
of the achievable out-of-sample performance in predicting $w$ from a local
$(u,v)$ stencil is already linear and physically grounded, captured by
the per-level regression of Section~\ref{sec:lr}, on both validation
domains; these are also the levels at which $w$ is still strongly
correlated with its value at the stencil level, whereas the performance
degrades as that correlation is lost with depth
(Section~\ref{sec:wcoherence}). At the two deepest levels tested, however, the FiLM
network shows a genuine advantage over this linear baseline on
the transition domain, while remaining at or below the linear baseline's
performance on Alboran, and providing, independently of that comparison, a
single model continuous in $z$ rather than one fit per level. This
suggests that the FiLM network does exploit non-linear structure away
from the reference level, but that this structure does not carry over to
Alboran, whose dynamics differ from those of the training domain
(Section~\ref{sec:domains}).
These conclusions are robust to the numerical uncertainty of the FiLM network: over
$274$ realizations differing only in the random seed, its out-of-sample
$R^2$ varies by $0.01$--$0.03$ on the transition domain and by up to $0.12$ on Alboran,
and each of the orderings above holds in between $82\%$ and $100\%$ of the
realizations. Whether
this depth-dependent gain strengthens with a wider stencil, more training
data, or additional validation domains is an open question this
comparison raises rather than settles.

Two further tests, reported in the Supporting Information, qualify these
results. The first (Text S1 in the Supporting Information) adds white
Gaussian noise to the input velocities of the transition domain. This
would mimic velocity measurement errors if the models were to be used
with empirical velocities as input. Both models, trained on noise-free data, are highly sensitive to it: a noise
with a standard deviation of $1\%$ of that of the velocities already
reduces the $R^2$ of the per-level regression by $12$--$14\%$, and at
$5\%$ both models fall below the performance of the domain mean almost
everywhere. The FiLM network is less sensitive, with a drop $1.8$ to
$4.6$ times smaller than that of the regression, but not robust either.
The second test (Text S2 in the Supporting Information) applies both
models, with no retraining, to $60$ weekly snapshots spanning about one year.
This would test whether the models perform well when set to predict $w$
at a time of the year different from the training one.
On the transition domain, below the stencil level, they fail from the end
of May to October, when $w$ near the surface is only weakly correlated
with $w$ below it; on Alboran the loss of performance is moderate. Models
trained on a summer snapshot have a low performance throughout the year,
because the near-surface stencil carries less information on $w$ already
on the training domain. The approach of this manuscript is therefore
suited to the season in which the near-surface flow is vertically coupled
to the flow below it, which is the season of the snapshot used here.

Natural extensions of the approach of this manuscript include incorporating the incompressibility
constraint of Equation~\ref{eq:continuity} directly into the training
objective as a physics-informed residual loss (evaluated via automatic
differentiation on the network output), in the spirit of hybrid
physics/ML approaches already explored for atmospheric
modeling \cite{Arcomano2022}, and comparing alternative mechanisms for injecting depth
into the network, for instance as an explicit, spatially-constant input
channel alongside $u$ and $v$, against the FiLM conditioning used here.

\section*{Conflict of Interest Statement}
The authors declare there are no conflicts of interest for this manuscript.

\section*{Acknowledgments}
This work has been supported by project PID2021-123352OB-C32 (LAMARCA)
funded by MICIU/AEI/10.13039/501100011033 and FEDER, `Una manera de hacer
Europa'. E.~H.-G.\ also acknowledges support by the Mar\'{\i}a de Maeztu
project CEX2021-001164-M funded by MICIU/AEI/10.13039/501100011033.
The authors used Claude (Anthropic), a large language model, to assist with
writing and revising the analysis and figure code, language editing, LaTeX
formatting and the checking of references. All content was reviewed by the
authors, who take full responsibility for the manuscript.

\bibliographystyle{apacite}
\bibliography{references}

\clearpage
\setcounter{equation}{0}
\setcounter{figure}{0}
\setcounter{table}{0}
\renewcommand{\theequation}{S\arabic{equation}}
\renewcommand{\thefigure}{S\arabic{figure}}
\renewcommand{\thetable}{S\arabic{table}}
\renewcommand{\theHequation}{S\arabic{equation}}
\renewcommand{\theHfigure}{S\arabic{figure}}
\renewcommand{\theHtable}{S\arabic{table}}

\begin{center}
{\LARGE Supporting Information}\\[1.5ex]
{\large for ``Estimating Vertical Ocean Velocity from Horizontal Currents: Linear and Depth-Conditional Neural-Network Approaches on a High-Resolution Regional Ocean Model''}
\end{center}
\vspace{1ex}

\section*{Introduction to the Supporting Information}
All results of the main text assume a perfectly known velocity field and
refer to a single winter snapshot of the simulation. Text S1 tests the
sensitivity of the two models of the main text, the per-level linear
regression and the depth-conditional FiLM network (FiLM), to errors in the
input horizontal velocities: white Gaussian noise is added to the input
$(u,v)$ of the transition domain, and both models are applied to the
noisy input exactly as trained on the noise-free training domain. For
the linear regression the expected loss of performance is derived
analytically and compared with the simulations, which provides an
independent check of the numerical procedure. Text S2 applies the two
models, trained on the winter snapshot, to $60$ snapshots spread over
about one year, relates the changes of their performance to the surface boundary
layer, the stratification and the vertical correlation of $w$, and
trains the two models again on a summer snapshot, to test whether the
behavior is symmetric between the two seasons. References to sections,
equations, figures and tables of the main text are given without the
prefix S.

\section*{Text S1. Sensitivity to errors in the input velocities}

\subsection*{S1.1 Noise experiment}
\label{sec:s_setup}

All results of the main text assume a perfectly known velocity field,
whereas any observed $(u,v)$ carries a measurement error. Adding noise to
the input velocities mimics such errors, as they would affect the models
if they were used with observed velocities as input.
\citeA{HeMahadevan2024} also tested input noise levels of $1$, $5$ and
$10\%$; here the sensitivity of both models to such an error is tested
by adding white Gaussian noise to the input
velocities of the transition domain (Section~2.3) at the stencil level,
before the stencils are extracted,
\begin{equation}
  u' = u + \alpha\,\sigma_u\,\xi_u, \qquad
  v' = v + \alpha\,\sigma_v\,\xi_v,
  \label{eq:s_noise}
\end{equation}
where $\xi_u$ and $\xi_v$ are independent standard normal variables,
drawn independently at every grid point, and $\sigma_u=0.165$~m\,s$^{-1}$
and $\sigma_v=0.121$~m\,s$^{-1}$ are the spatial standard deviations of
$u$ and $v$ over the transition domain. The dimensionless amplitude
$\alpha$ is the ratio between the standard deviation of the noise and
that of the velocity component to which it is added, the same for $u$
and $v$, so that $1/\alpha^2$ is the signal-to-noise ratio in variance.
Three amplitudes are used, $\alpha=1\%$, $5\%$ and $10\%$, i.e.\ a noise
standard deviation of $1.7$, $8.3$ and $16.5$~mm\,s$^{-1}$ on $u$ and
$1.2$, $6.1$ and $12.1$~mm\,s$^{-1}$ on $v$. The target is always the
unperturbed CROCO $w$.

Unlike He and Mahadevan, who trained their models on noisy data, both
models are used here exactly as trained on the noise-free training
domain: the test measures how a model fitted to perfect data responds to
an imperfect input, not how well it can learn to cope with one. FiLM
is evaluated with the final weights of $50$ of the realizations
of Section~2.5, each paired with an independent noise realization; the
per-level regression of Section~2.4 is evaluated on the same $50$ noisy
inputs, so that both models see identical perturbations. Performance is
measured with the $R^2$ of Equation~3 at the $16$ target levels.

\subsection*{S1.2 Effect of the noise on the linear regression}
\label{sec:s_lr}

For the per-level linear regression the effect of the noise can be
computed exactly. At a fixed level $k$, at grid point $n$ of the domain,
the regression of Equation~4 predicts
\begin{equation}
  \hat w_n = \sum_{p\in\mathcal N_r}\big(a_p\,u_{n,p} + b_p\,v_{n,p}\big) + c ,
\end{equation}
where the sum runs over the $49$ points $p$ of the stencil centered on
$n$, $a_p$ and $b_p$ are the coefficients $a^{(k)}_{\Delta j,\Delta i}$
and $b^{(k)}_{\Delta j,\Delta i}$, and $c$ is the intercept $c^{(k)}$.
Since the model is linear, the prediction with the noise of
Equation~\ref{eq:s_noise} is the noise-free one plus an additional term,
\begin{equation}
\begin{split}
  \hat w_n' &= \hat w_n + \delta_n, \\
  \delta_n &= \alpha\sum_{p\in\mathcal N_r}\big(a_p\,\sigma_u\,\xi_{u,n,p}
  + b_p\,\sigma_v\,\xi_{v,n,p}\big),
\end{split}
\end{equation}
i.e.\ the noise at each point of the stencil, weighted by the
corresponding coefficient.

Let $e_n=w_n-\hat w_n$ be the error of the model without noise and
$\mathrm{MSE}=\frac1N\sum_n e_n^2$ its mean squared error. With noise
the error at point $n$ becomes $e_n-\delta_n$, and expanding the square
gives
\begin{equation}
\begin{split}
  \mathrm{MSE}' &= \frac1N\sum_{n=1}^{N}\big(e_n-\delta_n\big)^2 \\
  &= \mathrm{MSE} - \frac2N\sum_{n=1}^{N} e_n\,\delta_n
  + \frac1N\sum_{n=1}^{N}\delta_n^2 .
\end{split}
\end{equation}
$\mathrm{MSE}'$ depends on the particular noise realization; we take its
expected value with respect to the $\xi$, bearing in mind that $w_n$,
$\hat w_n$ and $e_n$ are fixed quantities. The cross term has zero
expectation: by linearity
$\E\big[\frac1N\sum_n e_n\delta_n\big]=\frac1N\sum_n e_n\,\E[\delta_n]$,
and $\E[\delta_n]=0$ because every $\xi$ has zero mean. In other words,
a noise that is independent of the true $w$ cannot systematically
correct the error of the model. What remains is the quadratic term.
Writing $\delta_n^2$ as a double sum over the stencil points $p$ and $q$
and taking its expectation, all products of different noise variables
vanish, since $\E[\xi_{u,p}\xi_{u,q}]=0$ for $p\neq q$ and
$\E[\xi_{u,p}\xi_{v,q}]=0$ always, while the terms with $p=q$ give
$\E[\xi^2]=1$. Only the sum of the squared coefficients survives,
\begin{equation}
  \E[\delta_n^2] = \alpha^2\Big(\sigma_u^2\sum_{p}a_p^2
  + \sigma_v^2\sum_{p}b_p^2\Big),
\end{equation}
which is the same at every point $n$, since neither the coefficients nor
the noise amplitudes vary across the domain. Hence
$\E[\mathrm{MSE}']=\mathrm{MSE}+\E[\delta^2]$: the noise adds to the mean
squared error a non-negative amount, which vanishes only if all the
coefficients do.

Since the variance of $w$ at level $k$,
$\mathrm{Var}(w_k)=\frac1N\sum_n(w_n-\bar w)^2$, does not depend on the
noise, and $R^2=1-\mathrm{MSE}/\mathrm{Var}(w_k)$, the expected $R^2$
with noise is
\begin{equation}
  \E[R^{2\prime}] = R^2_0\,(1-\gamma),
  \qquad
  \gamma = \frac{\alpha^2\big(\sigma_u^2\sum_p a_p^2
  + \sigma_v^2\sum_p b_p^2\big)}{R^2_0\,\mathrm{Var}(w_k)},
  \label{eq:s_gamma}
\end{equation}
where $R^2_0$ is the noise-free value and $\gamma$ the relative loss.
The loss is quadratic in the amplitude $\alpha$, with a factor that
depends on the level only through the coefficients and the variance of
$w_k$. At the deeper levels, where the regression has smaller
coefficients and a prediction closer to the domain mean, the effect of
the noise is therefore smaller in absolute value.

The response of the linear regression to the noise is strong
(Table~\ref{tab:s_noise_lr}). Already at $\alpha=1\%$ it loses $0.109$ in
$R^2$ at $k=2$, i.e.\ $12\%$ of its noise-free value, and between $12\%$
and $14\%$ at the other three levels. At $\alpha=5\%$ its $R^2$ is
negative at every level: the error introduced by the noise exceeds the
variance of $w$ itself, so that the model errs more than a trivial
prediction equal to the domain mean $\bar w$ everywhere. The negative
values are largest near the surface, where the performance without noise
is highest, and smaller at depth. This follows from
Equation~\ref{eq:s_gamma}: a regression with lower performance has smaller
coefficients and hence a prediction closer to the mean, which is
unaffected by the noise, so that the drop scales with the noise-free
value $R^2_0$. A less negative $R^2$ at depth is therefore not a gain in
robustness, but the consequence of a prediction that was already closer
to the mean.

Equation~\ref{eq:s_gamma} provides an independent test of the whole
numerical procedure, from the generation of the noise and the
extraction of the stencils to the prediction and the computation of
$R^2$, since the mean over the $50$ noise realizations is an estimate of
$\E[R^{2\prime}]$. At $k=2$ the noise-free $R^2$ is $0.92199$ and the
expected value at $\alpha=1\%$ is $0.81313$, a drop of $0.10886$; because
of the $\alpha^2$ scaling, at $5\%$ and $10\%$ the expected drop is $25$
and $100$ times larger, $2.7215$ and $10.886$. The simulated values,
$0.8132\pm0.0010$, $-1.7967\pm0.0234$ and $-9.953\pm0.094$, differ from
the expected ones by $0.0001$, $0.003$ and $0.011$. Over all $16$ target
levels and three amplitudes the simulated values agree with
Equation~\ref{eq:s_gamma} to within $0.02$
(Table~\ref{tab:s_noise_lr}), a difference that never exceeds a quarter
of the standard deviation across noise realizations. The procedure is
therefore reliable, and it is applied unchanged to FiLM, for which no
such analytic prediction is available.

\begin{table}[!htb]
\centering
\begin{tabular}{rrccccc}
\toprule
$k$ & $z$ (m) & $R^2_0$ & $\alpha$ & $\gamma$ & $\E[R^{2\prime}]$ & simulated $R^{2\prime}$ (mean $\pm$ std) \\
\midrule
\multirow{3}{*}{2} & \multirow{3}{*}{$-11.2$} & \multirow{3}{*}{$0.9220$} & $1\%$ & $0.118$ & $\phantom{-}0.8131$ & $\phantom{-}0.8132\pm0.0010$ \\
 & &  & $5\%$ & $2.952$ & $-1.800$ & $-1.797\pm0.023$ \\
 & &  & $10\%$ & $11.807$ & $-9.964$ & $-9.953\pm0.094$ \\
\addlinespace
\multirow{3}{*}{7} & \multirow{3}{*}{$-37.1$} & \multirow{3}{*}{$0.8493$} & $1\%$ & $0.127$ & $\phantom{-}0.7412$ & $\phantom{-}0.7414\pm0.0012$ \\
 & &  & $5\%$ & $3.181$ & $-1.852$ & $-1.849\pm0.022$ \\
 & &  & $10\%$ & $12.724$ & $-9.956$ & $-9.942\pm0.086$ \\
\addlinespace
\multirow{3}{*}{12} & \multirow{3}{*}{$-71.1$} & \multirow{3}{*}{$0.3984$} & $1\%$ & $0.141$ & $\phantom{-}0.3421$ & $\phantom{-}0.3423\pm0.0008$ \\
 & &  & $5\%$ & $3.532$ & $-1.009$ & $-1.006\pm0.013$ \\
 & &  & $10\%$ & $14.129$ & $-5.231$ & $-5.222\pm0.050$ \\
\addlinespace
\multirow{3}{*}{17} & \multirow{3}{*}{$-116.6$} & \multirow{3}{*}{$0.2006$} & $1\%$ & $0.127$ & $\phantom{-}0.1752$ & $\phantom{-}0.1753\pm0.0005$ \\
 & &  & $5\%$ & $3.162$ & $-0.434$ & $-0.434\pm0.008$ \\
 & &  & $10\%$ & $12.649$ & $-2.337$ & $-2.338\pm0.029$ \\
\bottomrule
\end{tabular}
\vspace{10pt}
\caption{Per-level linear regression on the transition domain with white
  Gaussian noise of amplitude $\alpha$ added to the input velocities
  (Equation~\ref{eq:s_noise}). $R^2_0$: noise-free value (deterministic);
  $\gamma$ and $\E[R^{2\prime}]$: relative loss and expected value
  computed from the fitted coefficients (Equation~\ref{eq:s_gamma});
  simulated: mean $\pm$ standard deviation over $50$ noise
  realizations.}
\label{tab:s_noise_lr}
\end{table}

\subsection*{S1.3 Effect of the noise on FiLM}
\label{sec:s_film}

\begin{table}[!htb]
\centering
\begin{tabular}{rrcccc}
\toprule
$k$ & $z$ (m) & Noise-free & $\alpha=1\%$ & $\alpha=5\%$ & $\alpha=10\%$ \\
\midrule
 2 &  $-11.2$ & $0.869\pm0.012$ & $0.809\pm0.012$ & $-0.51\pm0.12$ & $-3.39\pm0.37$ \\
 7 &  $-37.1$ & $0.839\pm0.010$ & $0.782\pm0.007$ & $-0.27\pm0.08$ & $-2.37\pm0.28$ \\
12 &  $-71.1$ & $0.487\pm0.015$ & $0.462\pm0.015$ & $-0.03\pm0.08$ & $-1.19\pm0.29$ \\
17 & $-116.6$ & $0.237\pm0.025$ & $0.230\pm0.025$ & $\phantom{-}0.08\pm0.05$ & $-0.32\pm0.16$ \\
\bottomrule
\end{tabular}
\vspace{10pt}
\caption{Out-of-sample $R^2$ of FiLM on
  the transition domain with the same noise as in
  Table~\ref{tab:s_noise_lr}. Mean $\pm$ standard deviation over $50$
  realizations of FiLM (50 of the 274 realizations of Table~2, hence
  the small differences from the noise-free values there), each paired
  with its own noise realization.}
\label{tab:s_noise_film}
\end{table}

FiLM is less sensitive to the noise than the linear regression
(Table~\ref{tab:s_noise_film}): at the four levels and three amplitudes
of the tables its drop in $R^2$ is between $1.8$ and $4.6$ times smaller.
As a result, already at $\alpha=1\%$ FiLM outperforms the linear
regression at $k=7$, where without noise the regression led in $86\%$ of
the realizations (Section~3.4). At $\alpha=5\%$, however, the $R^2$ of
FiLM is also negative at every level except $k=17$.

For the linear regression the origin of this sensitivity can be read
from its coefficients. As shown in Section~3.3, it reconstructs $w$ through a
centered, antisymmetric stencil with the structure of the discrete
horizontal divergence, i.e.\ through differences between neighboring
grid points. Such differences are small for the velocity field, whose
values at neighboring grid points are close to each other, but not for a
noise that is uncorrelated from one grid point to the next, whose
difference between two points is as large as the noise itself. The noise
therefore dominates the estimated divergence even when it is small
compared with the velocity itself. Which features of the input FiLM
relies on is not known. Its lower sensitivity may be related to its
non-linearity, but the mechanism has not been investigated here.

Neither model, however, is robust to the noise, in contrast with the CNN
of \citeA{HeMahadevan2024}, whose squared correlation
decreased only from $0.76$ to $0.68$ at their $10\%$ noise level. The two
results are not directly comparable: their models were trained on noisy
data, their metric is a squared correlation over the upper $200$~m
rather than the level-by-level $R^2$ of Equation~3, and they do not state
the quantity to which their noise percentages refer. They suggest,
nonetheless, that robustness to noise is not an intrinsic property of a
convolutional network but is acquired when the noise is present during
training, at the cost of the smallest scales of the input.

\subsection*{S1.4 Summary}
\label{sec:s_noise_summary}

Both models are highly sensitive to errors in the input velocities when
they are trained on noise-free data: white noise with a standard
deviation of $1\%$ of that of the velocities already reduces the $R^2$ of
the linear regression by $12$--$14\%$, and at $5\%$ both models fall
below the performance of the domain mean almost everywhere. FiLM is
consistently less sensitive, with a drop $1.8$ to $4.6$ times smaller
than that of the linear regression, but this is not enough for it to be
robust. Applying either model to observed velocities will therefore
require training on data with a realistic error model.

\section*{Text S2. Temporal validation}
\label{sec:s_temporal}

The snapshot used in the main text, the average centered on 3 February
2011, 19:30~UTC, falls in winter (Section~2), and whether the results
carry over to other seasons is not tested there. The test below checks
whether the models perform well when set to predict $w$ at a time of the
year different from that of the training snapshot. Here the per-level
linear regression and FiLM, both trained on that snapshot, are applied,
with no retraining, to $60$
snapshots of the same simulation, one per week from 1 January 2011 to
12 February 2012. All of them are $3$-hour averages centered at
19:30~UTC, as the training snapshot, which is one of the $60$. The
per-level linear regression is that of Section~2.4 and is deterministic;
FiLM is the realization shown in Figure~9. Both are evaluated on
the two validation domains of Section~2.3, at the $16$ target levels,
with the $R^2$ of Equation~3. The fields of every snapshot are processed
exactly as in the main text; at the training snapshot this procedure
reproduces the $R^2$ of the linear regression in Table~1 and that of
this realization of FiLM.

\subsection*{S2.1 Seasonal cycle of the performance}
\label{sec:s_seasonal}

\begin{figure}[!htb]
  \centering
  \includegraphics{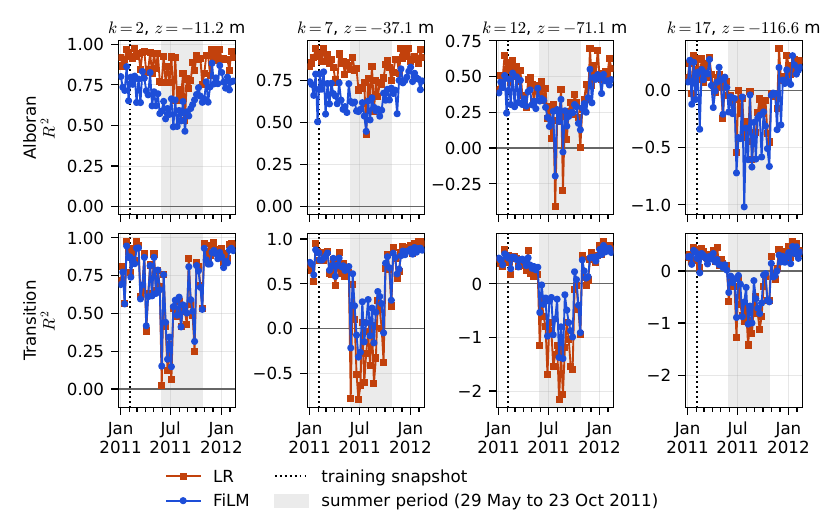}
  \caption{$R^2$ of the per-level linear regression (LR) and of FiLM
    (the realization of Figure~9), both trained on 3
    February 2011 (dotted line), at $60$ snapshots from 1 January 2011
    to 12 February 2012, on Alboran (top) and on the transition domain
    (bottom), at the four target levels of Table~2. Gray band: summer
    period.}
  \label{fig:s_r2_winter}
\end{figure}

The performance follows a clear seasonal cycle (Figure~\ref{fig:s_r2_winter}
and Table~\ref{tab:s_r2_periods}). On the transition domain it collapses
below the stencil level, for both models, between the end of May and
October 2011. At $k=12$ ($-71.1$~m) the $R^2$ of the linear regression
is negative at every one of the $22$ snapshots between 29 May and 23
October 2011, and at only one of the remaining $38$; that of FiLM
is negative at $20$ of these $22$ snapshots and at none of the remaining
$38$. We refer to these $22$ snapshots as the summer period. At $k=7$
($-37.1$~m) the $R^2$ is negative at $11$ of the $22$ summer snapshots
for the regression and at $8$ for FiLM, and at none of the
remaining $38$ for either model. The collapse starts and ends abruptly:
at $k=7$ the $R^2$ of the regression drops from $0.63$ on 22 May to
$-0.78$ on 29 May, and that of FiLM from $0.69$ to $-0.22$; on 30
October they are back to $0.90$ and $0.84$, from $0.25$ and $0.32$ a
week earlier. FiLM degrades less than the regression: in the
summer period its median $R^2$ is $0.24$ against $-0.02$ at $k=7$ and
$-0.51$ against $-1.09$ at $k=12$ (Table~\ref{tab:s_r2_periods}). On
Alboran the drop is moderate for both models: at $k=7$ the $R^2$ stays
between $0.43$ and $0.89$ for the regression and between $0.45$ and
$0.70$ for FiLM throughout the summer period.

\begin{table}[!htb]
\centering
\begin{tabular}{llcccc}
\toprule
 & & $k=2$ & $k=7$ & $k=12$ & $k=17$ \\
\midrule
\multirow{2}{*}{Alboran} & LR & $0.78\,|\,0.92$ & $0.75\,|\,0.87$ & $0.24\,|\,0.50$ & $-0.19\,|\,0.12$ \\
 & FiLM & $0.60\,|\,0.76$ & $0.60\,|\,0.73$ & $0.25\,|\,0.40$ & $-0.32\,|\,0.09$ \\
\addlinespace
\multirow{2}{*}{Transition} & LR & $0.51\,|\,0.89$ & $-0.02\,|\,0.81$ & $-1.09\,|\,0.45$ & $-0.66\,|\,0.30$ \\
 & FiLM & $0.52\,|\,0.85$ & $0.24\,|\,0.79$ & $-0.51\,|\,0.47$ & $-0.55\,|\,0.24$ \\
\bottomrule
\end{tabular}
\vspace{10pt}
\caption{Median $R^2$ of the models trained on 3 February 2011, over the
  $22$ snapshots of the summer period (29 May to 23 October 2011, left
  of the bar) and over the other $38$ snapshots (right of the bar).}
\label{tab:s_r2_periods}
\end{table}

\subsection*{S2.2 Boundary layer, stratification and vertical correlation of $w$}
\label{sec:s_hbl}

\begin{figure}[!htb]
  \centering
  \includegraphics{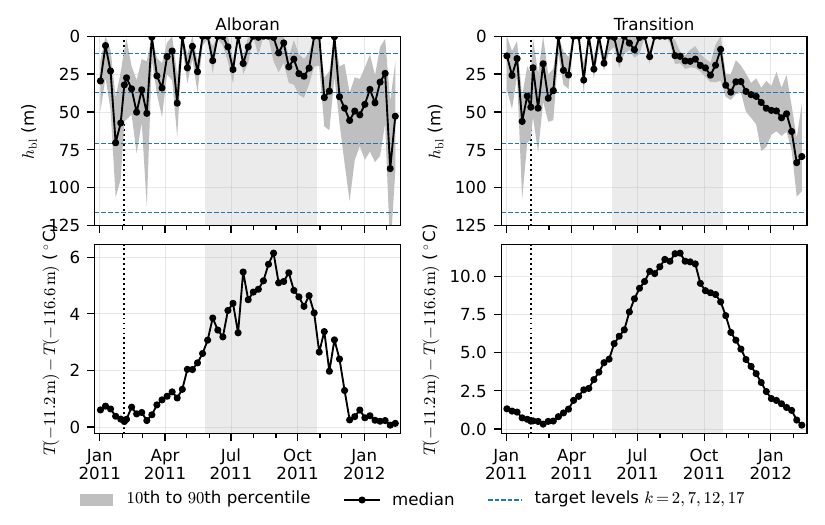}
  \caption{Top: depth $h_{\rm bl}$ of the surface boundary layer
    diagnosed by the model \protect\cite{Capo2021}, median (line) and $10$th to
    $90$th percentile range (shading) over the grid points of Alboran
    (left) and of the transition domain (right), at the $60$ snapshots
    of Figure~\ref{fig:s_r2_winter}. Dashed lines: target levels $k=2$,
    $7$, $12$ and $17$. Bottom: median over the domain of the
    temperature difference between the stencil level and the deepest
    target level. Dotted line: training snapshot; gray band: summer
    period.}
  \label{fig:s_hbl}
\end{figure}

\begin{figure}[!htb]
  \centering
  \includegraphics{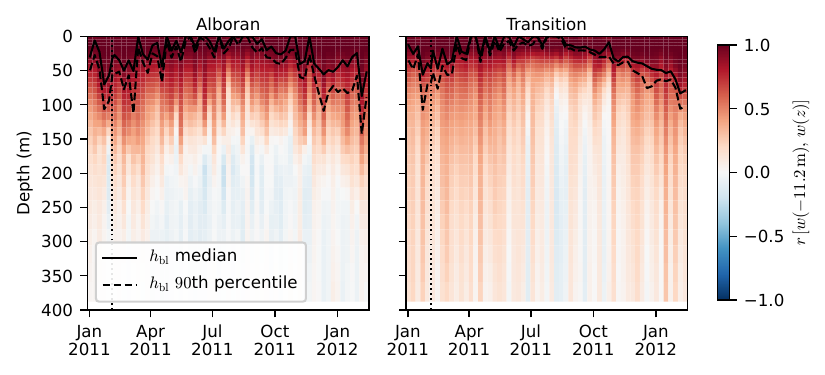}
  \caption{Correlation of $w$ at each depth with $w$ at the stencil level
    ($-11.2$~m), computed as in Figure~5, at the $60$ snapshots of
    Figure~\ref{fig:s_r2_winter}, on Alboran (left) and on the transition
    domain (right). Lines: median (solid) and $90$th percentile (dashed)
    of $h_{\rm bl}$. Dotted line: training snapshot.}
  \label{fig:s_wcorr}
\end{figure}

For every snapshot and domain we computed the depth $h_{\rm bl}$ of the
surface boundary layer (median and $10$th and $90$th percentiles over the
grid points, as in Section~3.2), the median temperature difference
between the stencil level and the deepest target level, as a measure of
the stratification, and the correlation of $w$ with its value at the
stencil level, as in Figure~5 (Figures~\ref{fig:s_hbl}
and~\ref{fig:s_wcorr}).

On the transition domain, in the summer period, the median $h_{\rm bl}$
is $10.9$~m against $34.1$~m in the rest of the year (medians over the
snapshots), the temperature difference $9.6^\circ$C against
$1.8^\circ$C, and the correlation of $w$ with its value at the stencil
level drops from $0.85$ to $0.66$ at $k=7$ and from $0.57$ to $0.32$ at
$k=12$. Over the $60$ snapshots, the rank (Spearman) correlation of the
$R^2$ of the regression at $k=7$ with the vertical correlation at the
same level is $0.96$, with the median $h_{\rm bl}$ $0.83$ and with the
temperature difference $-0.69$; at $k=12$ the three values are $0.93$,
$0.73$ and $-0.81$. For FiLM they are $0.96$, $0.83$ and $-0.70$
at $k=7$, and $0.92$, $0.71$ and $-0.82$ at $k=12$. A thin boundary
layer alone does not explain the collapse: on four snapshots in April
and May 2011 the median $h_{\rm bl}$ is zero, and yet at $k=7$ the $R^2$
is between $0.58$ and $0.68$ for the regression and between $0.65$ and
$0.71$ for FiLM. The quantity that best follows the performance is the
vertical correlation of $w$ itself, as anticipated in Section~3.2. On
Alboran the boundary layer is also thin in summer, with a median of
$5.7$~m against $34.5$~m, but the vertical correlation of $w$ changes
much less, $0.86$ against $0.92$ at $k=7$, consistent with the
correlation extending below the boundary layer there already in winter
(Section~3.2).

\subsection*{S2.3 Training on a summer snapshot}
\label{sec:s_summer}

\begin{figure}[!htb]
  \centering
  \includegraphics{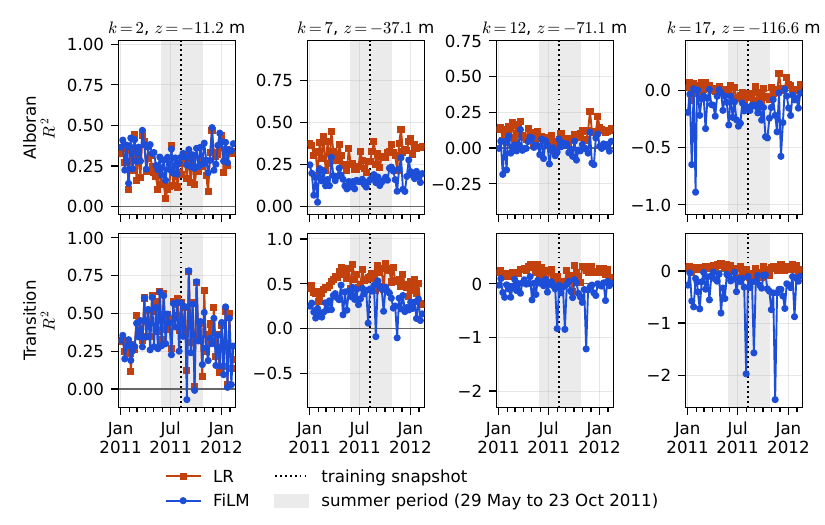}
  \caption{As Figure~\ref{fig:s_r2_winter}, for the models trained on 7
    August 2011 (dotted line). Each panel has the same vertical range as
    the corresponding panel of Figure~\ref{fig:s_r2_winter}.}
  \label{fig:s_r2_summer}
\end{figure}

To test whether the behavior is symmetric between the two seasons, both
models were trained again, with the same domain, procedure and random
seed, on the snapshot of 7 August 2011, 19:30~UTC, which lies in the
summer period, and evaluated at the same $60$ snapshots
(Figure~\ref{fig:s_r2_summer}).

The behavior is not symmetric: the models trained in summer do not
collapse in winter, but their performance is low throughout the year.
On the transition domain their median $R^2$ at $k=7$ is $0.57$ for the
regression and $0.42$ for FiLM in the summer period, and $0.49$
and $0.23$ in the rest of the year; at $k=12$ it is $0.14$ and $-0.04$,
and $0.22$ and $-0.04$. On Alboran it is $0.20$ and $0.28$ at $k=2$,
and $0.27$ and $0.15$ at $k=7$, in the summer period, against $0.31$
and $0.35$, and $0.33$ and $0.18$, in the rest of the year.

\begin{figure}[!htb]
  \centering
  \includegraphics{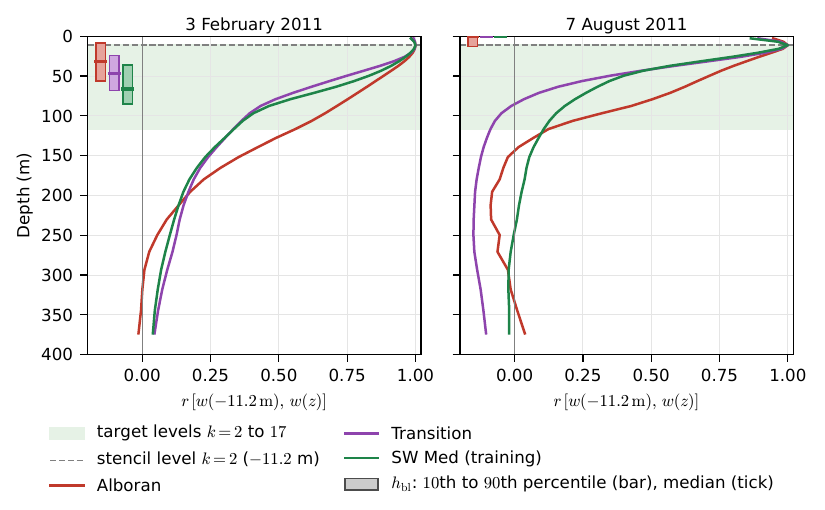}
  \caption{Correlation of $w$ at each depth with $w$ at the stencil
    level on 3 February (left, as in Figure~5) and on 7 August 2011
    (right), on the three domains. Vertical bars on the left: depth
    $h_{\rm bl}$ of the surface boundary layer, $10$th to $90$th
    percentile range over the grid points of each domain, with the
    median marked by a tick; on 7 August the median is below $1$~m on all
    three domains.}
  \label{fig:s_profiles}
\end{figure}

The reason lies in the training domain itself. On 7 August the $R^2$ of
both models on the data they were trained on is much lower than on 3
February (Table~\ref{tab:s_summer}): the near-surface stencil carries
less information on $w$ already there. On 3 February the correlation of
$w$ with its value at the stencil level on the training domain is
$0.91$, $0.63$ and $0.34$ at $k=7$, $12$ and $17$; on 7 August it is
$0.57$, $0.26$ and $0.11$, and even the $90$th percentile of
$h_{\rm bl}$ is zero (Figure~\ref{fig:s_profiles}). The training domain is
then in the same condition as the transition domain in the summer
period, and the relation learnt there does not apply to Alboran, where
$w$ remains coherent in depth in summer (correlation $0.80$ at $k=7$ on
7 August).

\begin{table}[!htb]
\centering
\begin{tabular}{lllcccc}
\toprule
 & & & $k=2$ & $k=7$ & $k=12$ & $k=17$ \\
\midrule
\multirow{4}{*}{Training} & \multirow{2}{*}{LR} & Feb & $0.898$ & $0.908$ & $0.628$ & $0.409$ \\
 & & Aug & $0.456$ & $0.581$ & $0.216$ & $0.075$ \\
 & \multirow{2}{*}{FiLM} & Feb & $0.880$ & $0.912$ & $0.729$ & $0.579$ \\
 & & Aug & $0.492$ & $0.687$ & $0.468$ & $0.350$ \\
\addlinespace
\multirow{4}{*}{Alboran} & \multirow{2}{*}{LR} & Feb & $0.924$ & $0.890$ & $0.546$ & $0.183$ \\
 & & Aug & $0.197$ & $0.267$ & $0.060$ & $-0.059$ \\
 & \multirow{2}{*}{FiLM} & Feb & $0.756$ & $0.696$ & $0.401$ & $0.143$ \\
 & & Aug & $0.304$ & $0.159$ & $0.063$ & $-0.148$ \\
\addlinespace
\multirow{4}{*}{Transition} & \multirow{2}{*}{LR} & Feb & $0.922$ & $0.849$ & $0.398$ & $0.201$ \\
 & & Aug & $0.573$ & $0.584$ & $-0.051$ & $-0.108$ \\
 & \multirow{2}{*}{FiLM} & Feb & $0.874$ & $0.844$ & $0.495$ & $0.240$ \\
 & & Aug & $0.569$ & $0.441$ & $-0.026$ & $-0.105$ \\
\bottomrule
\end{tabular}
\vspace{10pt}
\caption{$R^2$ of the models trained on 3 February (Feb) and on 7 August
  2011 (Aug), each evaluated at its own training snapshot: on the
  South-West Mediterranean training domain, i.e.\ on the data the model
  was fitted on, and on the two validation domains. FiLM is a
  single realization, with the same random seed in the two cases.}
\label{tab:s_summer}
\end{table}

\subsection*{S2.4 Summary}
\label{sec:s_temporal_summary}

Both models, trained on a winter snapshot, fail on the transition domain
below the stencil level in the stratified season, from the end of May to
October, and recover from the end of October; on Alboran the loss of
performance is moderate. The performance of both models is limited by
how strongly $w$ near the surface is coupled to $w$ below it, both where
a model is applied and where it is trained.

Where the coupling is weak in the domain where a model is applied, even
a model trained under strong coupling fails. This is the case of the
models trained in winter on the transition domain in summer, whose $R^2$
follows the vertical correlation of $w$ with a rank correlation of
$0.96$ at $k=7$, more closely than the depth of the boundary layer or
the stratification.

Where the coupling is weak in the domain where a model is trained, the
model learns a weak relation and its performance is low throughout the
year. On the
training domain the correlation of $w$ at $k=7$ with its value at the
stencil level is $0.91$ on 3 February 2011, and $0.57$ and $0.71$ on 7
August and on 4 September 2011, a second summer snapshot on which the
models were trained with the same procedure. The $R^2$ of FiLM on
its own training data at $k=7$ is correspondingly $0.91$ when trained in
February, against $0.69$ and $0.73$ when trained on the two summer
snapshots. At $k=12$ the difference is larger: the correlation is
$0.63$, against $0.26$ and $0.30$, and the $R^2$ of FiLM $0.73$,
against $0.47$ and $0.51$.

The approach
of the main text is therefore suited to the season in which the
near-surface flow is vertically coupled to the flow below it, which is
the season of the snapshot used there.

\end{document}